\documentclass[rpreprint,3p,10pt,sort&compress, colorlinks,linkcolor=blue, citecolor=blue]{elsarticlex}

\usepackage[utf8]{inputenc}

\usepackage{amssymb}
\usepackage{amsmath,mathrsfs}
\usepackage{color}
\usepackage{placeins}

\usepackage{graphicx}
\usepackage{subcaption}
\usepackage{amsmath}
\usepackage{pgfplots}
\usepackage{tikz}
\pgfplotsset{compat=1.18}
\usepgfplotslibrary{groupplots}
\usetikzlibrary{pgfplots.groupplots}

\usepgfplotslibrary{groupplots}
\usetikzlibrary{pgfplots.groupplots}

\font\LARGEbsf=cmssdc10 scaled 2100      

\usepackage{siunitx}

\usepackage{stfloats}

\usepackage{bm}

\usepackage{graphicx}

\usepackage{caption}
\usepackage{subcaption}

\usepackage{multicol}

\usepackage{multirow}

\usepackage{mhchem}

\usepackage[toc,page]{appendix}

\usepackage{bm}

\usepackage[pagewise]{lineno}

\usepackage{float}
\usepackage[makeroom]{cancel}

\usepackage[colorlinks,linkcolor=blue,anchorcolor=green,citecolor=blue]{hyperref}

\usepackage{natbib}
\definecolor{saffron}{HTML}{FF9933}
\definecolor{brickred}{HTML}{F96302}
\usepackage{siunitx}

\begin{document}
	
	\begin{frontmatter} 
		\title{Hierarchical multiscale fracture modeling of carbon-nitride nanosheet reinforced composites by combining cohesive phase-field and molecular dynamics}
		
		\author[els,*]{Qinghua Zhang}
		\ead{qinghua.zhang@outlook.com}
		
		\author[els]{Navid Valizadeh}
		
		\author[rvs]{Mingpeng Liu}

		\author[els]{Xiaoying Zhuang}
		\author[els,*]{Bohayra Mortazavi}
		\ead{bohayra.mortazavi@gmail.com}

		
		\cortext[cor1]{Corresponding author}
		
		\address[els]{Chair of Computational Science and Simulation Technology, Department of Mathematics and Physics, Leibniz University Hannover,  Welfengarten 1A, 30167 Hannover, Germany}
		\address[rvs]{School of Qilu Transportation, Shandong University, Jinan, 250002, China}
		
		\begin{abstract}
			Understanding the fracture mechanisms in composite materials across scales, from nano- to micro-scales, is essential for an in-depth understanding of the reinforcement mechanisms and designing the next generation of lightweight, high-strength composites. However, conventional methods struggle to model the complex fracture behavior of nanocomposites, particularly at the fiber-matrix interface. The phase-field regularized cohesive fracture model has proven to be effective in simulating crack initiation, branching, and propagation; however, capturing the cohesive fracture strength at smaller scales remains a significant challenge. This study introduces a novel approach that combines an energy-based star-convex decomposition cohesive phase-field fracture model with molecular dynamics simulations to explore the thickness dependency of nanocomposite mechanical properties. The proposed framework enables hierarchical modeling of carbon-nitride nanosheet-reinforced composites' mechanical and fracture behaviors. The developed model could elucidate complex fracture processes across different scales and highlight critical scaling effects. This methodology provides an efficient solution for uncovering hierarchical fracture mechanisms in reinforced nanocomposites, offering valuable insights into their fracture behavior and strengthening mechanisms.
		\end{abstract}
		
		\begin{keyword}
			Hierarchical multiscale modeling; Molecular dynamics; Cohesive phase-field fracture; Scaling effects;  Carbon-nitride nanosheet; Atomistic-continuum modeling.
		\end{keyword}
		
	\end{frontmatter}
	
	\section{Introduction}
	\label{Introduction}
	Since the first mechanical exfoliation of graphene \cite{novoselov2004electric,geim2007rise,schedin2007detection,neto2009electronic} in 2004 and the synthesis of g-$\rm{C_{3}N_{4}}$ \cite{liu2015metal} in 2009, two-dimensional materials such as graphene (GN) and carbon-nitride ($\text{C}_{x}\text{N}_{y}$) nanosheets with few-layer thicknesses have attracted significant attraction in both academy and industry. Among these materials, graphene plays a pioneering role in the field of two-dimensional materials due to its simple lattice structure and remarkable properties, including thermal conductivity \cite{shahil2012graphene,balandin2008superior}, mechanical strength \cite{lee2008measurement}, and electronic characteristics \cite{neto2009electronic}. These outstanding physical properties have enabled graphene to be extensively attractive as a functional component to enhance properties of composites across various scales, such as nanoscale 
	microscale, 
	 and even the macroscale \cite{dong2021physicochemical}. Recently, several investigations \cite{hatam2020thermal,rajabpour2019carbon} have highlighted that experimentally synthesized $\rm{C_{3}N}$ \cite{mahmood2016two} exhibits comparable mechanical properties 
	and thermal conductivity similar to graphene, 
	while also possessing a more stable structure compared to other carbon-nitride monolayers, 
	These studies highlight the remarkable properties and practical applications of both GN and $\text{C}_{x}\text{N}_{y}$ across a wide range of industrial sectors.
	
	The key aspect that leads to the different mechanical 
	and thermal transport 
	properties of composite systems and heterostructures at different scales is related to the dimensions of the fillers. At the nanoscale or in nanocomposites, the surface area between the fillers and the matrix per volume is much higher compared to the microcomposites. In order to simulate and capture the corresponding interfacial properties, it is crucial to first investigate the atomic interactions between the fillers and the matrix, which are usually polymeric systems. This is usually realized by incorporating and carefully analyzing the non-banding \textit{van der Waals} interactions.
	At the nanoscale, \textit{van der Waals} interactions have been widely utilized to model cohesive interactions among atoms. However, the \textit{van der Waals}-driven multiscale cohesive fracture behavior of $\text{C}_{x}\text{N}_{y}$-reinforced composites remain underexplored in the existing research literature. The complexity of fracture criteria, along with challenges in mitigating information loss during cross-scale modeling, has created significant gaps in understanding the multiscale fracture response of both GN and $\text{C}_{x}\text{N}_{y}$, particularly at the atomic and continuum levels.	Moreover, the investigation of the fracture behavior of the pristine carbon-nitride monolayer, which has a thickness of only a few angstroms (\AA), presents a significant challenge in the experimental setting. To overcome these challenges, molecular dynamics (MD) simulations combined with multiscale modeling provide a promising approach for investigating the mechanical and fracture behaviors of carbon-nitride materials. In the context of multiscale modeling, various methods offer frameworks to thoroughly investigate complex physical behaviors across different scales. These approaches are generally classified into three main categories: hierarchical, semi-concurrent, and concurrent methods \cite{talebi2014computational, budarapu2019multiscale}.
	
	Among the aforementioned multiscale methods, the hierarchical multiscale method is capable of transferring and bridging information from lower to upper scales. However, it can not perform a reverse transfer. This method's high computational efficiency makes it particularly suitable for hierarchically modeling new two-dimensional materials, such as investigating the mechanical properties of graphene/borophene (GN/BN) heterostructures 
	and graphene-reinforced composites. 
	In contrast, the semi-concurrent multiscale method focuses on bi-directional information exchange between lower and upper scales. A typical example is the $\rm{FE^2}$ method \cite{feyel1999multiscale,feyel2000fe2} which exemplifies this approach. However, the $\rm{FE^2}$ method faces significant computational challenges due to the parallel computations required at both the micro and macro scales. This limitation restricts its application primarily to simple, two-dimensional academic examples. As a result, it is less effective in accurately representing the complex characteristics of composite materials, such as GN or $\text{C}_{x}\text{N}_{y}$ platelet reinforcements. Additionally, the $\rm{FE^2}$ method suffers from instability in cases of material softening, which can result from damage or other degradation mechanisms \cite{budarapu2019multiscale}.
	
	At the macroscale, classical continuum mechanics effectively models the fracture behavior of materials using several degradation functions to degrade material stiffness in order to realize the damage and fracture of materials 
	\cite{danesh2021comparative,danesh2022free}.
	However, it exhibits limited capability to describe the initiation and propagation of cracks arising from atomic dislocations at the nanoscale. This limitation arises from the difference in modeling mechanisms between continuum mechanics and molecular dynamics. In continuum mechanics, partial differential equations (PDEs) govern the equilibrium conditions within a continuous medium and are typically solved using discretization techniques, like finite elements method \cite{wriggers2008nonlinear,miehe2010phase,miehe2010thermodynamically}, meshfree free method \cite{rabczuk2019extended,ren2016dual,ren2017dual}, etc. However, at the nanoscale, material deformations are governed by atomic dislocations. 
	Additionally, accurately capturing the fracture process presents substantial challenges for conventional continuum mechanics, where the de-bonding of individual atoms at the crack tips leads to the formation of microscopic voids (damage).
	
	This study aims to explore the thickness-dependent fracture mechanisms of GN and other prominent $\text{C}_{x}\text{N}_{y}$ nanosheets reinforced polymer composites, by developing a state-of-the-art hierarchical multiscale method. This method starts from the natural properties of materials, circumventing the challenges typically associated with the natural discontinuity of cracks on the RVE boundary by a displacement jump. This enables a more focused exploration of the material properties. However, the complexity and discontinuity of material fracture modeling require diverse evolution criteria. To address this issue, the phase-field fracture model provides a flexible and elegant solution. For modeling crack propagation within RVEs at the microscale, the cohesive phase-field fracture model \cite{wu2017unified,rezaei2022anisotropic} is preferred rather than the displacement jump-based cohesive zone models \cite{rezaei2017prediction,tarafder2020finite}. This preference stems from the strong capability of the phase-field model to simulate crack initiation, propagation, coalescence, branching \cite{zhuang2022phase,zhou2019phase}, and phase transformations \cite{danesh2022thermodynamically,danesh2021nonlocal}. Furthermore, the cohesive zone model suffers from discontinuities within the fracture zone and significantly relies on predefined duplicate nodes in the fracture zone. To address these limitations, this study employs a hybrid approach that combines a cohesive phase-field model and molecular dynamics method. The combined approach aims to reveal the crack patterns of graphene and carbon-nitride as well as the fracture behavior of their reinforced composites over multiple scales.
	
	On account of the above issues, the content of this study is organized as follows. In \hyperref[Sec2: MD and CPF]{{Section 2}}, details of the classical molecular dynamics modeling developed to acquire the mechanical and fracture responses of the considered monolayers (referred to as fibers), cohesive interfaces, and polymer matrix at the atomic scale are provided. Next, the homogenization theory developed to facilitate the bridge between the nanoscale and microscale is given. Subsequently, the isotropic/anisotropic cohesive phase-field fracture theory is presented at the microscale. \hyperref[Sec3: Model validation]{{Section 3}} employs two numerical benchmark tests and experimental data to validate the material model at the microscale. \hyperref[Sec4: Results and discussion]{{Section 4}} presents an in-depth discussion of the simulation results across different scales. Finally, \hyperref[Sec5: Conclusion]{{Section 5}} provides concluding remarks and outlook.
	
	\section{Computational method}\label{Modelling}	
	\label{Sec2: MD and CPF}
	\subsection{Nanoscale, mechanics/cohesive fracture modeling}
	\label{nanoscale_2_1}
	\subsubsection{Mechanical properties of nanosheets and polymer matrix}
	
\begin{figure*}[!h]
		\centering
		\includegraphics[width=18cm]{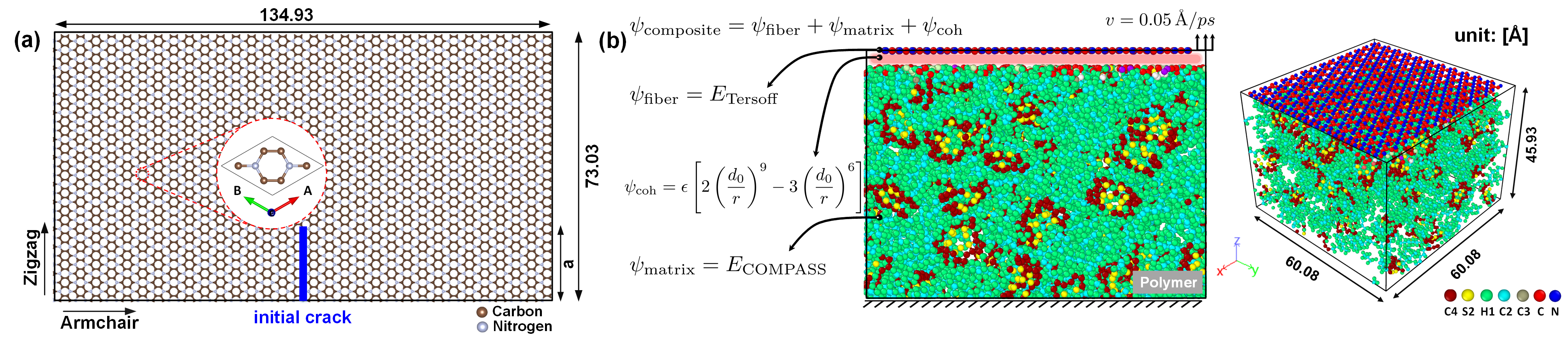}
		\caption{A $\rm{C_{3}N}$ monolayer with an initial crack length of $a$ \text{\AA} and containing 3840 atoms (in the absence of a notch) is visualized using \textit{VESTA} \cite{momma2011vesta}. An enlarged view highlights the monolayer’s lattice structure. (b) A schematic illustration of the cohesive zone model is shown by employing a $\rm{C_{3}N}$ monolayer under a hybrid empirical interatomic potential that governs the fiber, interface, and matrix by $\psi_{{\rm{Tersoff}}}$, $\psi_{\rm{coh}}$, and $\psi_{\rm{COMPASS}}$, respectively.}
		\label{fig1}
	\end{figure*}

	At the nanoscale, molecular dynamics (MD) simulations are performed by employing the open-source package \textit{LAMMPS} \cite{plimpton1995fast} to calculate the homogenized elastic modulus $E$, Poisson's ratio $\rm{\nu}$, critical energy release rate $G_c$, and cohesive fracture strength $\sigma _{0}$ between the fibers ($\rm{GN}$, $\text{C}_{x}\text{N}_{y}$) and matrix (P3HT). The atomic structure of the representative two-dimensional carbon-nitride ($\rm{C_{3}N}$) nanosheet is depicted in \hyperref[fig1]{Fig.1 (a)}. The brown and grey atoms in \hyperref[fig1]{Fig.1 (a)} denote carbon and nitrogen atoms, respectively. Analogously, the corresponding atoms for the cohesive zone model are depicted in \hyperref[fig1]{Fig.1 (b)}.
	
	The accuracy of MD simulations relies heavily on the appropriate selection of empirical interatomic potentials and the corresponding parameter sets. 
	For instance, to eliminate non-physical strain hardening of graphene at high strain levels and achieve a more accurate stress-strain curve aligned with experimental results, the cutoff distance for the C-C atomic bond is modified within the range of  $r_{c}=0.20$ nm to $r_{c}=0.21$ nm.
	Consequently, the optimized \textit{Tersoff} potential with a modified cutoff is employed to describe C-C bond interactions in uniaxial tensile tests of graphene, as proposed by \citet{lindsay2010optimized} and \citet{sevik2012influence}. Furthermore, \citet{lipomi2011stretchable} proved that the optimized \textit{Tersoff} potential is the most accurate for describing the mechanical and thermal transport interactions of $sp^2$ carbon atoms in molecular dynamics simulations. The next step is to establish periodic boundary conditions in both the armchair and zigzag directions for the RVE using \textit{LAMMPS}. With the boundary conditions and the optimized \textit{Tersoff} potential in place, uniaxial tensile tests for GN and $\text{C}_{x}\text{N}_{y}$ are conducted in the armchair direction. The constant timestep and loading rate for the two-dimensional reinforcements are set as 0.5 femtoseconds (\textit{fs}) and $6\times10^{9}s^{-1}$, respectively, using the \textit{metal} unit system in \textit{LAMMPS}. Subsequently, a Gaussian-distributed velocity is initialized for the specimen prior to the simulation, followed by a heating process from 100 K to 300 K under the NPT ensemble for 30 picoseconds (\textit{ps}) to eliminate internal residual stress. In this context, the structures of the reinforcements are relaxed. The temperature of the specimen is then equilibrated for 20 \textit{ps} to reach the target temperature (300 K) before the uniaxial tensile tests. Thereafter, uniaxial tensile tests are performed on the reinforcements to calculate and output the homogenized stress per picosecond. The fundamental principle for homogenizing local atomic stress under periodic boundary conditions is derived from the virial theorem \cite{clausius1870xvi,maxwell1870reciprocal}, as expressed in \hyperref[1]{Eq.\eqref{1}} below:
	
	\begin{equation}
		{\bm{\sigma} _{\rm{virial}}} = \frac{1}{V}\sum\limits_{i \in V}^{{N_A}} {\Big[ {\frac{1}{2}\sum\limits_{i \ne j}  {{\bm{r}_{ij}} \otimes {\bm{F}_{ij}} - {m_i} \,  {{\dot {\bm{u}}}_i} \otimes {{\dot {\bm{u}}}_i}} } \Big]} \quad {\rm{with}} \quad {\bm{F}_{ij}} = \frac{{\partial {\psi _{\rm{nano}}}\left( {r_{ij}} \right)}}{{\partial {r_{ij}}}}\frac{{\bm{r}_{ij}}}{ {r_{ij}}}.
		\label{1}
	\end{equation}
	
	\noindent Here, the homogenized virial stress (${\bm{\sigma} _{\rm{virial}}}$) of the RVE enables bridging the atomic scale to the continuum scale by averaging the local stress in molecular dynamics simulations \cite{chen2006mathematical}, which represents the equivalent mechanical response of the target monolayer (RVE). In \hyperref[1]{Eq.\eqref{1}}, $V$ and $N_A$ denote the volume of the RVE and the number of atoms inside the specimen, respectively. ${{\psi _{\rm{nano}}}}$ represents the empirical interatomic potential of the specimen at the nanoscale with the actual distance $r_{ij}$ between atoms $i$ and $j$, expressed as ${r_{ij}}:= ||\bm{r}_{ij}|| = || \bm{x}_{i} -  \bm{x}_{j} ||$. $\bm{x}_i$ and $\bm{x}_j$ denote the actual position vectors of atoms $i$ and $j$, respectively, and $\bm{r}_{ij}/r_{ij}$ represents the unit vector of $\bm{r}_{ij}$. ${{\psi _{\rm{nano}}}}$ could consist of the potentials from the fiber ($\psi_{{\rm{Tersoff}}}$), cohesive interface ($\psi_{\rm{coh}}$), and matrix ($\psi_{\rm{COMPASS}}$). Here, $m_i$ and $\dot{\bm{u}}_i$ denote the mass and velocity of atom $i$, respectively, and the operator $\otimes$ represents the dyadic product of two vectors to generate a second-order tensor, as shown in \hyperref[1]{Eq.\eqref{1}}.
	
	To obtain the critical energy release rates of pristine GN and $\text{C}_{x}\text{N}_{y}$ nanosheets, two notches of different lengths ($10$ \text{\AA} and $35$ \text{\AA}) are designed in the middle of the specimen along the zigzag direction (see \hyperref[fig1]{Fig.1 (a)}). These specimens are then subjected to a uniaxial tensile test under external loading at 300 K, with periodic boundary conditions applied along the armchair and zigzag directions. To avoid size-dependent effects and strong fluctuations in homogenized data measurements at the nanoscale, a sufficiently large RVE (approximately $135 \, \text{\AA}  \times 73 \, \text{\AA} $) is employed, with initialized single notches of $1 \, \text{\AA}  \times 10 \, \text{\AA} $ and $1 \, \text{\AA} \times 35 \, \text{\AA} $. The size-dependent effects of homogenized stress and strain on specimen size were investigated in our previous study 
	. With the homogenized data in hand, the critical energy release rates ($G_{c}$) of the monolayers are calculated by referring to \cite{zhang2014atomistic,bao2018molecular} as follows:
	
	\begin{equation}
		G_{c} =  - \frac{{dW}}{{2 \, {\delta _{th}} \, da}} = \frac{{W_{a}-W_{a + \Delta a}}}{{2 \, {\delta _{th}} \, da}} \quad {\rm{with}} \quad W = \int\limits_t {\sum\limits_{i = 1}^{{N_A}} {{\bm{F}_i}(t) \cdot {\bm{v}_i}} \, dt},
		\label{2}
	\end{equation}
	
	\noindent where $W_a$ and $W_{a+\Delta a}$ denote the fracture energy of the monolayers with predefined initial crack lengths of $a$ \text{\AA} and $a+\Delta a$ $\text{\AA}$, respectively. Therefore, the external work is expressed as an integration of the whole time for the virtual work, which is induced by the external force vector of atoms $\bm{F}_i$ and the corresponding velocity $\bm{v}_i$ vector of the atom $i$ over all atoms $N_A$. $\delta _{th}$ represents the geometrical thickness of target monolayers. In \hyperref[2]{Eq.\eqref{2}}, the value $2$ stems from the two fractured \textit{lines} or \textit{surfaces} created by the silting of materials while employing two- or three-dimensional MD simulations, respectively. Here, the geometrical thickness of graphene and other carbon-nitride monolayers are defined as $3.35$ $\text{\AA}$ 
	and $3.40$ $\text{\AA}$ 
	, respectively. Therefore, $G_c$ can be evaluated via \hyperref[2]{Eq.\eqref{2}} by the anticipated homogenized result in molecular dynamics simulations. 
	
	Following the investigation of the mechanical properties, uniaxial tensile tests are conducted to evaluate the matrix's mechanical properties in the $X$, $Y$, and $Z$ directions for a cubic specimen with a length of 60.1 $\text{Å}$. For brevity, a simplified formulation of the molecular dynamics simulation process is provided here. For more details, the reader is referred to our previous work 
	. In this regard, the simulation time increment is set to 0.3 \textit{fs}. The boundary conditions for the specimen are defined as periodic (\textit{PPP}) along three axes. A Gaussian-distributed initial velocity is then applied to the specimen. Before tensile loading, the system is equilibrated at 500 K under the NPT ensemble for 50 \textit{ps} to eliminate residual stress resulting from non-equilibrated atomic fluctuations. Subsequently, the structure is cooled from 500 K to 300 K under the NPT\footnote{The isothermal-isobaric ensemble is a statistical mechanical system that maintains constant temperature (T) and pressure (P) throughout the process.} ensemble for 50 \textit{ps}. To obtain a stable temperature (300 K) for the specimen during the tensile process, further equilibration is performed using the NPT ensemble at 300 K for 50 \textit{ps}. Finally, the specimen is subjected to tensile loading along the three axes separately, during which homogenized normal stress is printed every two picoseconds.

	\subsubsection{Specific choice of potential expressions for cohesive model}
	As depicted in \hyperref[fig1]{Fig.1(b)}, the cohesive fracture strength $\sigma_0$ is calculated from the normal traction of a single monolayer of the matrix. Initially, the optimization of the specimen ($60.1 \,\text{\AA} \times 60.1 \, \text{\AA} \times 45.9 \, \text{\AA}$) is performed using the conjugate gradient (CG) algorithm with a relatively smaller timestep of 0.1 \textit{fs} in a \textit{real} unit system in \textit{LAMMPS}, to eliminate any residual stress from the non-equilibrium state. The boundaries are defined as \textit{PPS}, which apply periodic boundary conditions in the $ X $ and $ Y $ directions and a shrink-wrapped response in the $ Z$ direction. Afterwards, the \textit{Tersoff} potential \cite{lindsay2010optimized}, \textit{van der Waals}-induced cohesive potential, and the second-generation (\textit{COMPASS} \cite{sun1998compass}) coupled novel hybrid potentials are used to formulate the atomic response for the fibers, interfaces, and the matrix, respectively.
	
	\begin{equation}
		{\psi _{\rm{composite}}} = {\psi _{\rm{Tersoff}}}\left( r, \theta \right) + {\psi _{\rm{coh}}}\left( r \right) + {\psi _{\rm{COMPASS}}}\left( r, b, \theta , \vartheta , \chi \right),
		\label{3}
	\end{equation}
	
	\noindent where the potential of the composite $\psi _{\rm{composite}}$ is formulated by the actual distance $r$ between two atoms, atomic bond length $b$, bond-bending angle $\theta$, torsion angle $\vartheta$, and out-of-plane angle $\chi$, see \cite{budarapu2019multiscale,sun1998compass}. The empirical interatomic \textit{Tersoff} potential $\psi _{{\rm{Tersoff}}}$ is defined as:
	
	\begin{equation}
		{\psi _{{\rm{Tersoff}}}} = f_{C} \, ( {\underbrace {\mathcal{A} \, {e^{ - {\lambda _1} \, r}}}_{{f_R}} - \underbrace {{b_{ij}} \, \mathcal{B} \, {e^{ - {\lambda _{2}} \, r}}}_{{f_A}}}) \quad {\rm{with}} \quad {b_{ij}} = {\left( {1 + \beta ^{m} \, \varsigma _{ij}^{m}} \right)^{ - 1/2m}},
		\label{4}
	\end{equation}
	
	\noindent where $f_{R}$ and $f_{A}$ denote the repulsive and attractive pairwise terms, respectively, which are expressed by exponential Morse-like functions. These two terms are induced by their corresponding energies $\mathcal{A}$ and $\mathcal{B}$, respectively. $f_{C}$ denotes a smooth spherical cutoff function with a predefined cutoff-based quantity to govern the first nearest-neighbor interaction. $b_{ij}$ formulates the atomic interactions induced by three-body bond angle $\theta _{ijk}$ among atoms $i$, $j$, and $k$ as well as their coordinates. Detailed expansion of $\varsigma _{ij}^{m}$ and material parameters (like $\beta$, $m$, etc.) in \hyperref[4]{Eq.\eqref{4}} are referred to \hyperref[AppendixA]{Appendix A} and literature works \cite{lindsay2010optimized,sevik2011characterization}. 
	
	Due to the complicated extensions in $\psi_{\rm{COMPASS}}$, a more detailed expression is not provided here but is referred to in the contributions by \citet{sun1998compass} and \citet{budarapu2019multiscale}. Regarding the cohesive interaction between the fiber and matrix in \hyperref[3]{Eq.\eqref{3}}, the Lennard-Jones 9-6 potential is employed to describe the atomic interactions. This specific choice is made because of the \textit{softer} interaction in the interfacial region between the fibers and matrix, compared to the \textit{stronger} repulsive interactions typically characterized by the Lennard-Jones 12-6 potential, see the comparisons in \hyperref[fig11]{Appendix B}. Moreover, the Lennard-Jones 12-6 potential is more suitable for first-generation force fields like \textit{AMBER} \cite{pearlman1995amber}, \textit{CHARMM} \cite{mackerell1998all}, and \textit{CVFF} \cite{dauber1988structure}. Hence, the cohesive interaction driven by the non-bonded Lennard-Jones 9-6 potential \cite{sharma2019molecular, budarapu2019multiscale} is expressed as follows:
	
	\begin{equation}
		{\psi _{\rm{coh}}} = \left\{ {\begin{array}{*{20}{l}}
				{\epsilon \left[ {{{2\left( {\frac{{{E _0}}}{r}} \right)}^9} - {{3\left( {\frac{{{E _0}}}{r}} \right)}^6}} \right],} & {r \le {r_c}}\\
				0, & {r > {r_c}}
		\end{array}} \right.,
		\label{5}
	\end{equation}
	
	\noindent where $ \epsilon $, $ E_0 $, and $ r_c $ represent the initial cohesive energy, the actual distance among atoms at zero cohesive energy, and the predefined cutoff in the cohesive interaction, respectively. Here, the cutoff is defined as $ r_c = 12.0 \, \text{\AA} $ between fibers and the matrix. The detailed contribution of cohesive energy between the matrix and fiber for specific atoms is shown in \hyperref[AppendixB]{Appendix B}. Subsequently, the traction-separation simulation is conducted at 300 K with the hybrid potential represented by \hyperref[3]{Eq.\eqref{3}}. The fibers ($\rm{GN}$, $\text{C}_x\text{N}_y$) are then prescribed with a velocity of 0.05 $\text{\AA}/{fs}$ along the $ Z $-direction. The specimen below 5 $\text{\AA}$ is prescribed as rigid, as shown in \hyperref[fig1]{Fig.1(b)}. The simulation is conducted at 300 K under the NVT\footnote{The NVT ensemble is a statistical model used to study material properties when the number of particles (N) and the volume (V) are fixed, while the temperature (T) varies around an equilibrium value.} ensemble to equilibrate the composite for 100 \textit{ps}. Next, all the local atomic stresses along the $ Z $-direction ($ \sigma_{zz} $) are homogenized using \hyperref[1]{Eq.\eqref{1}}, and the homogenized stress is output every picosecond. Subsequently, the representative cohesive strength at the microscale is obtained by filtering out the maximum stress via $ \sigma_0 = {\rm Max} \, (\sigma_{zz}) $. Following the hierarchical multiscale modeling concept for graphene/borophene heterostructures proposed by 
	, the homogenized material parameters $ \bm{\zeta}_p = \left[ E, \, \nu, \, G_c, \, \sigma_0 \right] $ at the nanoscale are calculated and employed in the microstructure to investigate the cohesive phase-field fracture response of RVEs.
	
	\subsection{Microscale, phase-field regularized cohesive fracture modeling}
	\subsubsection{Governing equations} 
\begin{figure*}[!h]
		\centering
		\includegraphics[width=14cm]{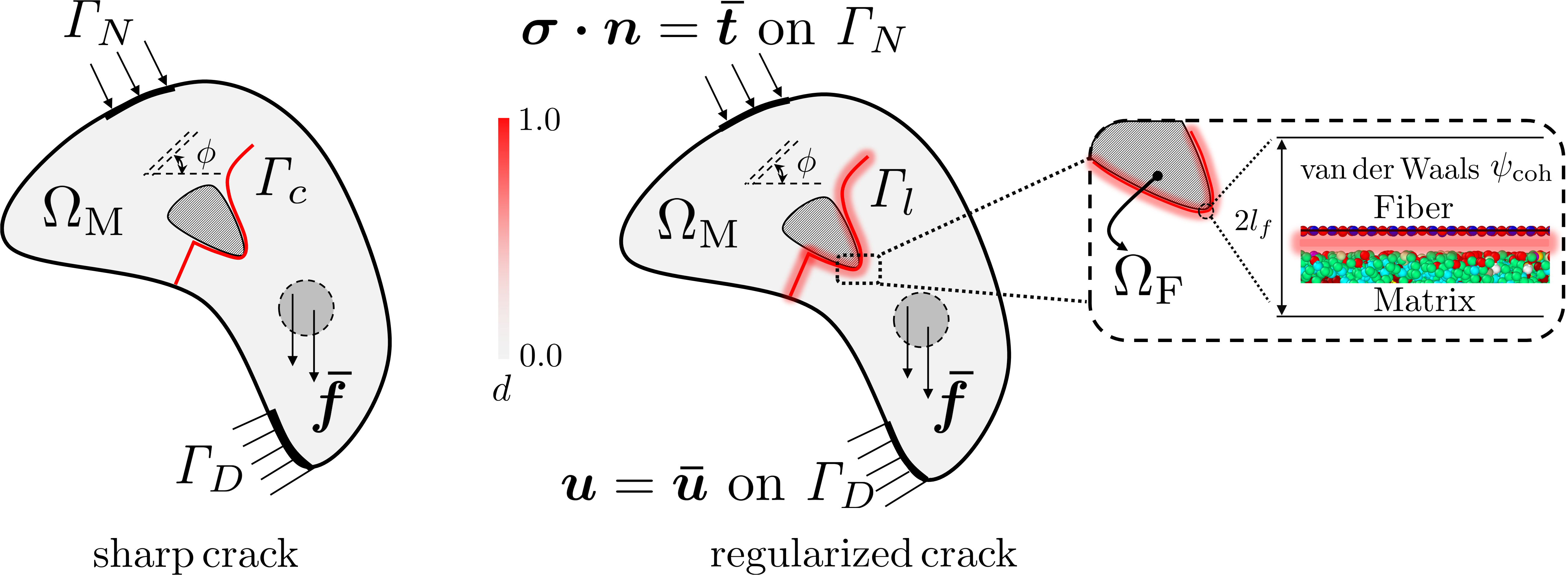}
		\caption{A schematic illustration depicts an arbitrary solid matrix domain $ \Omega_{\rm{M}} $ and a fiber domain $ \Omega_{\rm{F}} $ with a sharp crack $ \mathit{\Gamma}_c $ and a phase-field regularized crack $ \mathit{\Gamma}_l $. Here, $ l_f $ denotes the diffused length of the regularized crack band. $ \bar{\bm{f}} $ and $ \bar{\bm{t}} $ represent the body force in $ \Omega $ and the traction on the Neumann boundary $ \mathit{\Gamma}_N $, respectively. The anisotropic fiber orientation is expressed by a predefined angle $ \phi $.}
		\label{fig111}
	\end{figure*}

	At the microscale, based on the Ginzburg-Landau equation, the cohesive phase-field fracture model (CPF) is employed to distinguish between the two phases of the solid: the fracture phase ($d = 1$) and the undamaged phase ($d = 0$). Consider an elastic continuum body $ \Omega \subset \text{I\!R}^n $ with dimensional space $ n = 2, 3 $ bounded by $ \mathit{\Gamma} $, where $ \mathit{\Gamma} \subset \Omega $. Moreover, $ \mathit{\Gamma} $ can be decomposed into two disjoint sets: $ \mathit{\Gamma}_N $, where Neumann boundary conditions are prescribed, and $ \mathit{\Gamma}_D $, where Dirichlet boundary conditions are applied, as shown in \hyperref[fig111]{Fig.2}. The mechanical response of the material points $ \bm{x} \in \Omega $ at time $ t $ is formulated as $ \bm{u}(\bm{x}, t) $ for the displacement field. Owing to the employed P3HT, the polymer exhibits brittleness below the glass transition temperature. A small strain tensor $ \bm{\varepsilon} $ is used to describe the kinematics of the polymer matrix at the microscale in this study, as follows:
	\begin{equation}
		\boldsymbol{\varepsilon}  =  {\rm{sym}}\left[ {\nabla _s} \boldsymbol{u} \right]: = \frac{1}{2}\left( { \nabla \boldsymbol{u} + \nabla \boldsymbol{ {u}}^{\rm{T}}} \right),
		\label{6}
	\end{equation}
	
	\noindent where $\nabla$ represents the  gradient operator. Based on the MD simulation results for the matrix (\hyperref[fig6]{Fig.7 (a)}) and fiber (\hyperref[fig5]{Fig.6}), the phase-field fracture behavior is assumed to be isotropic for the matrix and anisotropic for the fiber, respectively.
	
	\FloatBarrier
	
	\begin{equation}
		\mathit{\Gamma}_{c} = \underbrace {\int_{\mathcal{S}} \, dA }_{{\rm{sharp \, crack}}} \approx \mathit{\Gamma}_{l} = \underbrace {\int_\Omega  {\gamma \left( {d,\nabla d} \right) \, dV} }_{{\rm{regularized \, crack}}}
		\label{7}
	\end{equation} 
	
	\noindent The main concept of the phase-field method is to smear the sharp crack $ \mathit{\Gamma}_c $ into a diffused damage band $ \mathit{\Gamma}_l $, where the damage field $ d \, (\bm{x}, t) $ is distributed such that $ 0 \le d \, (\bm{x}, t) \le 1 $, with $ \bm{x} \in \Omega $. The phase-field variable $ d $ is determined by minimizing the diffused crack topology, expressed as $d \, (\bm{x},t) = {\rm{Arg}} \{ {\mathop {\inf }\limits_d {\mathit{\Gamma} _l} \, (d)} \}$. Following the image segmentation theory proposed by \citet{braides1998approximation}, the crack density function is used to reformulate Griffith's fracture energy $ W_{\rm{frac}} = G_c \, \mathit{\Gamma}_l $ into a volumetric integral form, as expressed in \hyperref[7]{Eq.\eqref{7}}.
	
	In phase-field fracture modeling, the sharp crack surface topology $\mathit{\Gamma} _{c}$ of the RVE is regularized using the cohesive phase-field crack surface density function $\gamma$ (see \cite{wu2018robust}), resulting in $\mathit{\Gamma} _{l}$. The latter consists of the regularized cracks in the fiber $\mathit{\Gamma} _{\rm{F}}$ and in the matrix $\mathit{\Gamma} _{\rm {M}}$, such that $\mathit{\Gamma} _{l}  = {\mathit{\Gamma} _{\rm{F}}} \cup {\mathit{\Gamma} _{\rm{M}}}$. The fracture density function $\gamma$ was initially introduced by \citet{ambrosio1990approximation} for the normalized segmentation in Mumford-Shah variation \cite{mumford1989optimal}. This has been consistently proven by \citet{braides1998approximation} that the sharp crack surface $\mathit{\Gamma}_c$ satisfies the $\mathit{\Gamma}$-convergence theorem by employing the regularized crack surface $\mathit{\Gamma}_l$ as $\mathit{\Gamma}_c = \mathop {\lim }\limits_{l_f \to 0} {\mathit{\Gamma}_{l}} \, (d)$. Consequently, the fracture energy $W_{{\rm{frac}}}$ is expressed as:
	\begin{equation}
		W_{{\rm{frac}}} \left( d, \nabla d \right) = \int_{\Omega} {{G_c ^{i}} \,\, {\gamma ^{i}}\left( {d,\nabla d} \right)} \, dV \quad {\rm{with}} \quad {{\gamma ^{i=\rm{M}}}\left( {d,\nabla d} \right) = \frac{1}{{{c_0}}}\left( {\frac{1}{{{l_f}}}\omega \left( d \right) + {l_f} \, \nabla d \cdot \nabla d} \right)},
		\label{8}
	\end{equation}
	\noindent where $i = \{\text{{F, M}}\}$, when $i={\text{F}}$ or $i={\text{M}}$, the corresponding quantities of fiber and matrix are employed, respectively. The normalization constant $c_0$ is calculated by \hyperref[8-1]{Eq.\eqref{8-1}} as ${c_0}=\pi$.
	\begin{equation}
		{c_0} = 4\int_0^1 {\omega (s) \, ds} \quad {\rm{with}} \quad\omega (d) = 2 \, d - {d^2}
		\label{8-1}
	\end{equation}
	$ \omega (d) $ denotes a crack geometric function that characterizes the phase-field crack evolution. Regarding the fiber, considering the anisotropic mechanical behavior of GN and $ \text{C}_x\text{N}_y $ from prior studies 
	along with simulation findings in \hyperref[fig5]{Fig.6}, the crack density function $ \gamma^{i = \rm{F}} $ for the fiber is expressed using a second-order structure tensor $ \mathbf{A} $, incorporating a predefined anisotropic fracture angle $ \phi $.
	
	\begin{equation}
		{{\gamma ^{i=\rm{F}}}\left( {d,\nabla d; \phi} \right) = \frac{1}{{{c_0}}}\left( {\frac{1}{{{l_f}}}\omega \left( d \right) + {l_f} \, \nabla d \cdot {\mathbf{A}} \cdot \nabla d} \right)}
		\label{9}
	\end{equation} 
	\noindent where ${\mathbf{A}}=\bm{I}+\alpha \, \bm{a} \otimes \bm{a}$ with $\bm{a} = {[ {\begin{array}{*{20}{c}}
				{\cos \left( {\phi} \right)}&{\sin \left( {\phi} \right)}
		\end{array}} ]^{\rm{T}}}$. $\bm{I}$ and $\alpha$ denote the identity matrix and penalization parameter for anisotropic orientation, respectively.  
	\begin{equation}
		{\mathbf{A}}= \bm{I} + \alpha \, \left[ {\begin{array}{*{20}{c}}
				{\cos^2 {{\left( {\phi} \right)}}}&{\cos \left( {\phi} \right)\sin \left( {\phi} \right)}\\
				{\cos \left( {\phi} \right)\sin \left( {\phi} \right)}&{\sin ^2 {{\left( {\phi} \right)}}}
		\end{array}} \right]
		\label{10}
	\end{equation}
	
	\noindent In this study, the nanosheets are considered to exhibit an anisotropic fracture response in the $ X $-$ Y $ plane based on the observed lattice structure and fracture trajectories in \hyperref[fig1]{Fig.1} and \hyperref[fig5]{Fig.6}, respectively.
	
	\subsubsection{Energy density functions} 
	The total density function $ \psi $, describing the mechanical response within a domain, is expressed as the sum of its internal part $ \psi_{\rm{in}} $ and external part $ \psi_{\rm{ext}} $, as follows:
	
	\begin{equation}
		\psi  = {\psi _{{\rm{in}}} \left(\bm{\varepsilon}, d, \nabla d; \phi \right)} - {\psi _{{\rm{ext}}} \left(\bm{u} \right)},
		\label{11}
	\end{equation}
	\begin{equation}
		\psi _{{\rm{ext}}} \left(\bm{u} \right) = \bar{\bm{f}} \cdot \bm{u} + \bar{\bm{t}} \cdot \bm{u}.
		\label{12}
	\end{equation}
	
	\noindent The internal part represents the material stored energy density function $\psi_{{\rm{in}}}$, which comprises the bulk energy density $\psi _{\rm{bulk}}$ and the fracture energy density $\psi _{\rm{frac}}$ as follows:
	\begin{equation}
		\psi _{{\rm{in}}} \left(\bm{\varepsilon}, d, \nabla d; \phi \right) = {\psi _{{\rm{bulk}}}}\left( {\bm{\varepsilon} ,d} \right) + {\psi _{{\rm{frac}}}}\left( {d,\nabla d; \phi } \right).
		\label{13}
	\end{equation}
	
	\noindent $ \psi_{\rm{bulk}} $ is decomposed into two components, corresponding to tension and compression, referred to as the positive stored strain energy density $ \psi_{e}^{+} $ and the negative energy density $ \psi_{e}^{-} $, respectively. These components are defined in terms of the degradation function $ g(d) $ and the damage threshold $ \psi_{\rm{th}} $, following the formulations proposed by \citet{miehe2010phase}. Therefore, $ \psi_{\rm{bulk}} $ is expressed as:
	\begin{equation}
		\psi _{{\rm{bulk}}} \left( \bm{\varepsilon}, d \right) = g\left( d \right)\psi _e^ + \left( \bm{\varepsilon}  \right) + \psi _e^ - \left( \bm{\varepsilon}  \right).
		\label{14}
	\end{equation}
	The fracture energy density function $\psi_{{\rm{frac}}}$ is formulated as:
	\begin{equation}
		\psi _{{\rm{frac}}} \left( d, \nabla d;\phi \right) = [1 - g\left( d \right)] \, {\psi _{th}} + \frac{{2 \, {\psi _{th}}}}{\zeta }{l_f} \, \gamma^i \left( {d,\nabla d ; \phi } \right).
		\label{15}
	\end{equation}
	
	\noindent The degradation function interpolates between the broken state ($ d = 1 $) and the undamaged state ($ d = 0 $), satisfying the conditions $ g(0) = 1 $, $ g(1) = 0 $, $ g'(d) \le 0 $, and $ g'(1) = 0 $. The latter condition ensures that the fractured band does not continuously increase once the material begins to fracture. $ \psi_{\rm{th}} $ represents the damage threshold, and $ \zeta $ controls the behavior in the post-critical range following crack initiation. Therefore, the degradation function is expressed as:
	
	\begin{equation}
		g(d) = \frac{{{{\left( {1 - d} \right)}^p}}}{{{{\left( {1 - d} \right)}^p} + {a_1} \, d\left( {1 + {a_2} \, d + {a_3} \, {d^2}} \right)}}.
		\label{16}
	\end{equation}
	
	\noindent 
	In \hyperref[16]{Eq.\ref{16}}, $a_1$, $a_2$, and $a_3$ are used to characterize the phase-field regularized cohesive traction-separation law \cite{nguyen2018modeling}, which governs material softening. The parameter $a_1$ is determined at the initial state ($d=0$) of the internal energy density $ \psi_{\rm{in}} $ with Griffith's critical energy release rate $ G_c $ (see \hyperref[8]{Eq.\eqref{8}}), while assuming an isotropic case without considering asymmetric decomposition into tension and compression. The brief derivation is presented as follows:
	
	\begin{equation}
		\frac{{\partial {\psi _{{\rm{in}}}}}}{{\partial d}} = g'(d) \, {\psi _e} + \frac{{{G_c}}}{{{c_0}}}\left( {\frac{{\omega '(d)}}{{{l_f}}}} \right) = 0 \quad 
		\xrightarrow[]{d=0}
		\quad 
		{a_1} = \frac{{4 \, {l_{ch}}}}{{\pi \, {l_f}}} \quad {\rm{with}} \quad {l_{ch}} = \frac{{E \, G_{c}}}{{\sigma _0^2}}.
		\label{17}
	\end{equation}
	
	\noindent Here, Irvin's internal length $l_{ch}$ in \hyperref[17]{Eq.\eqref{17}} governs the size of fractures during crack evolution, incorporating Young's modulus $E$, the critical energy release rate $G_c$, and the material fracture strength $\sigma_0$. The detailed derivations to obtain $g'(d)$ and $a_1$ can be found in \hyperref[AppendixC]{Appendix C}. For the softening response, this study primarily introduces the linear softening \cite{wu2017unified} and the experimentally-based softening models \cite{cornelissen1986experimental} for the cohesive fracture behavior.
	\begin{itemize}
		\item Remark 1: Linear softening: $p = 2, \, {a_2} =  - \frac{1}{2}, \, {a_3} = 0$
		\begin{equation}
			\sigma \left( w \right) = {\sigma_0} \, {\rm{Max}} \left( {1 - \frac{{\sigma_{0}}}{2 \, G_{c}}w,0} \right)
			\label{18}
		\end{equation}
		\item  Remark 2: Cornelissen softening: $p = 2, \, {a_2} = 1.3868, \, {a_3} = 0.6567$
		\begin{equation}
			\sigma \left( w \right) = {\sigma_0}\left[ {\left( {1 + \xi _1^3 \, {r_w^3}} \right){e^{ - {\xi _2} \, r_w}} - r_w \, \left( {1 + \xi _1^3} \right){e^{ - {\xi _2}}}} \right]
			\label{18-2}
		\end{equation}
	\end{itemize}
	
	The stress $\sigma$ versus separation $w$ relationship induced by material softening is described using the parameters $\xi_1 = 3.0$, $\xi_2 = 6.93$, and a normalized crack opening $r_w := w / w_c$, as explained by \citet{nguyen2018modeling}. Here, $\sigma_0$ represents the failure strength at $w = 0$, and the critical separation displacement $w_c$ is determined by $w_c = 2 \, G_c / \sigma_0$. Specifically, for linear softening, $w_c$ is calculated as $w_c = 5.1361 \, G_c / \sigma_0$. Further insights into these parameters can be found in \citet{wu2017unified}.
	
	To model a more realistic crack trajectory in platelet-included composites under tensile and compression loading scenarios, a star-convex energy-decomposed model \cite{vicentini2024energy} is employed by incorporating an additional parameter $\gamma^* \geq -1$ to flexibly calibrate the ratio $\bm{\tau}$ between compression and tension stresses, $\bm{\tau} = {\bm{\sigma}_e^-}/{\bm{\sigma}_e^+}$. The star-convex parameter is chosen as $\gamma^* = 3$ to govern precise crack evolution. This approach signifies that the initialization of crack surfaces is realized by degrading not only the conventional positive (tensile) term but also the negative (compression) mixed term, as illustrated in \hyperref[19]{Eq.\eqref{19}}.
	
	\begin{equation}
		{\rm{Matrix:}} 
		\left\{
		\begin{aligned}
			{\psi _e^ {+} (\bm{\varepsilon})} &  = {\frac{\kappa }{2} \left[\left\langle {{\rm{tr}}(\bm{ \varepsilon}  )} \right\rangle _ + ^2 - \gamma ^*  \left\langle {{\rm{tr}} (\bm{ \varepsilon} ) } \right\rangle _ - ^2  \right] + \mu \, \left( {\bm{\varepsilon} ': \bm{\varepsilon} '} \right)} 
			\\
			{\psi _e^ {-} (\bm{\varepsilon})} & =  (1+\gamma ^ *) {\frac{\kappa }{2}\left\langle {{\rm{tr}} (\bm{ \varepsilon}  )} \right\rangle _ - ^2}
		\end{aligned}
		\right.\\
		\label{19}
	\end{equation}
	
	\begin{equation}
		{\rm{Fiber:}}  \left \{ \begin{aligned}
			{\psi _e^ {\pm} (\bm{\varepsilon})} &  = \dfrac{\lambda }{2}\left\langle {{\rm{tr}} (\bm{ \varepsilon} ) } \right\rangle _ \pm ^2 + \mu \, \left( \bm{{\varepsilon} _ {\pm}} : \bm{{\varepsilon} _ {\pm}} \right)
		\end{aligned}
		\right.\\
		\label{20}
	\end{equation}
	
	From the above decompositions and cohesive softening equations, the degradation function $g(d)$ plays a crucial role in quantifying the degradation of elastic energy as damage evolves. The asymmetric decomposition into tensile and compression behaviors can be defined based on various criteria, such as strain \cite{miehe2010phase}, stress \cite{fei2021double}, and energy \cite{hudobivnik2022adaptive} considerations in brittle fracture scenarios. In \hyperref[19]{Eq.\eqref{19}} and \hyperref[20]{Eq.\eqref{20}}, the Macaulay bracket operator is defined as ${\left\langle \bullet \right\rangle _ \pm } = ({{\bullet \pm | \bullet |}})/2$ to avoid the negative eigenvalues from the decomposition in \hyperref[21]{Eq.\eqref{21}}. Regarding the positive and negative strain tensors $\bm{\varepsilon}_{\pm}$, they are obtained through the spectral decomposition of the strain tensor $\bm{\varepsilon}$ as follows.
	
	\begin{equation}
		{\bm{\varepsilon} = \bm{\varepsilon}_{+} + \bm{\varepsilon}_{-}}
		\quad
		{\rm{with}}
		\quad
		{\bm{\varepsilon} _ \pm } := \sum\limits_{i = 1}^3 {{{\left\langle {{{\varepsilon} _i}} \right\rangle }_ \pm } \, {\bm{n}_i} \otimes {\bm{n}_i}},
		\label{21}
	\end{equation}
	
	\noindent where ${\varepsilon}_i$ denotes the principal strain (eigenvalues) and $\bm{n}_i$ represents the principal strain direction (eigenvectors) of the strain tensor. $\bm{\varepsilon}'$ represents the deviatoric strain, expressed as $\bm{\varepsilon}' =\bm{\varepsilon}  - {1/3} \, {\rm{tr}(\bm{\varepsilon})} \, \bm{I}$. $\kappa$ denotes the bulk modulus which can be formulated by the Lam\'e constants as $\kappa  = \lambda  + 2 \mu /n$. The  \hyperref[19]{Eq.\eqref{19}} indicates that crack evolution is influenced by positive and negative volumetric expansion, deviatoric strains, and distortions in material shape. 
	
	Compared to the computational cost of computing the deviatoric strain $\bm{\varepsilon}'$, the calculation of eigenvalues and eigenvectors in \hyperref[21]{Eq.\eqref{21}} to obtain $\bm{\varepsilon}_{\pm}$ is significantly higher. This approach leads to a highly nonlinear stress-strain constitutive model. Therefore, the deviatoric strain decomposition method is preferred for formulating the mechanical behavior in the matrix region. Additionally, the decomposition of stored energy based on deviatoric strain has proven effective in inhibiting crack propagation in composite and masonry structures \cite{haghighat2023efficient}. Furthermore, because the matrix constitutes the majority of the materials in the RVE and most cracks occur within the matrix rather than the fiber, it is practical to apply spectral decomposition only for modeling the fiber fractures. This selective approach helps enhance computational efficiency.
	\subsubsection{Dissipative potential function} 
	In this section, the rate-independent potential function proposed by \citet{miehe2010phase} and \citet{aldakheel2018phase} is employed due to its natural dissipation characteristics associated with crack propagation.
	
	\begin{equation}
		\mathit{\Pi} ^*( {\dot {\boldsymbol{u}},\dot d}) = \int_{\Omega}  {W( {\dot {\boldsymbol{u}},\dot d} ) \, dV}  - \int_{\Omega}  {\bar {\boldsymbol{f}} \cdot \dot {\boldsymbol{u}} \, dV}  - \int_{\partial {\Omega}} {\bar {\boldsymbol{t}} \cdot \dot {\boldsymbol{u}} \, d\mathit{\Gamma}}
		\label{22}
	\end{equation}
	\noindent where, the potential density function ${W( {\dot {\boldsymbol{u}},\dot d} )}$ is defined as:
	
	\begin{equation}
		W( {\dot {\boldsymbol{u}},\dot d} ) = \frac{d}{{dt}}\psi \left( {\dot {\boldsymbol{u}}} \right) + \mathcal{D} \, ( {\dot d}) = \frac{d}{{dt}}\psi  + \underbrace{\left({\delta _d}\psi \right)  \, \dot d - \frac{1}{{2 \, \eta }}\left\langle {{\delta _d}\psi }\right\rangle _ + ^2}_{\mathcal{D}}.
		\label{23}
	\end{equation}
	
	\noindent Here, $\mathcal{D}$ represents the dissipation potential that governs a mixed variational principle. Therefore, by minimizing \hyperref[22]{Eq.\eqref{22}} yields the evolutions of displacement and phase-field as $[{\dot {\boldsymbol{u}},\dot d} ] = {\rm{Arg}}\{ {\mathop {\inf }\limits_{\bm{\dot u}, \, \dot d} \left( {\mathit{\Pi} ^*} \right)} \}$. Further variational derivations of the rate-independent potential in \hyperref[23]{Eq.\eqref{23}} with respect to $\bm{\dot u}$ and $\dot{d}$ leads to the balance of linear momentum, fracture driving force $\mathcal{F}^f$, and dissipative resistance force $\mathcal{R}^f$ as follows:
	
	\begin{itemize}
		\item Displacement field
	\end{itemize}
	\begin{equation}
		{{\delta _{\bm{\dot u}}}W = \bm{0}},
		\quad
		{{\rm{Div}}\left( {{\partial _{\bm{\varepsilon}} }\psi } \right) + \bar {\bm{f}} = \bm{0}}
		\label{24}
	\end{equation}
	
	\begin{itemize}
		\item Phase field
	\end{itemize}
	\begin{equation}
		{\delta _{\dot d}}W = {0}, 
		\quad
		- ({\partial _d}\psi  - {\rm{Div}}({\partial _{\nabla d}}\psi )) + {\delta _d}\psi = {0}, \quad
		\mathcal{F}^f - \mathcal{R}^f = \underbrace {{\partial _d}\psi  - {\rm{Div}}({\partial _{\nabla d}}\psi )}_{{\delta _d}\psi }
		\label{25}
	\end{equation}
	
	\begin{itemize}
		\item Phase-field evolution
	\end{itemize}
	\begin{equation}
		\frac{\partial W}{\partial ({\delta _d}\psi )} = 0, 
		\quad
		\dot d - \frac{1}{\eta }{\left\langle {{\delta _d}\psi } \right\rangle _ + } = 0
		\label{26}
	\end{equation}
	
	\noindent Consequently, the objective is to derive ${{\delta _d}\psi }$. By utilizing $\rm{\psi _{bulk}}$ from \hyperref[14]{Eq.\eqref{14}}, one can obtain:
	
	\begin{equation}
		{\partial _d}\psi  = {\partial _d}{\psi _{{\rm{bulk}}}} + {\partial _d}{\psi _{{\rm{frac}}}} \quad {\rm{with}} \quad {\psi _{{\rm{frac}}}} = (1 - g(d)) \, {\psi _{th}} + \frac{{2 \, {\psi _{th}}}}{\zeta }{l_f} \, \gamma^i \left( {d,\nabla d; \phi } \right),
		\label{27}
	\end{equation}
	
	\begin{equation}
		{\partial _d}\psi  = g'(d)\left[ {\psi _e^ +  - {\psi _{th}}} \right] + \frac{{2 \, {\psi _{th}}}}{\zeta }\frac{1}{{{c_0} \, {l_f}}} \omega '(d).
		\label{28}
	\end{equation}
	
	\noindent The partial derivative of the anisotropic energy density function with respect to $\nabla d$ is expressed as:
	
	\begin{equation}
		{\partial _{\nabla d}}\psi  = \cancelto{\bm{0}} {{\partial _{\nabla d}}{\psi _{{\rm{bulk}}}}} + {\partial _{\nabla d}}{\psi _{{\rm{frac}}}} = \frac{{2 \, {\psi _{th}}}}{{{c_0} \, \zeta }}2 \, l_f^2 \, \nabla d \cdot {\mathbf{A}}.
		\label{29}
	\end{equation}
	Subsequently, ${\delta _d}\psi$ can be obtained by using the divergent operator to \hyperref[29]{Eq.\eqref{29}} as ${{\rm{Div}}({\partial _{\nabla d}}\psi )}$ in the following:
	
	\begin{equation}
		{\delta _d}\psi  = g'(d)\left[ {\psi _e^ +  - {\psi _{th}}} \right] + \frac{{2 \, {\psi _{th}}}}{{{c_0} \, \zeta }}\left[ {\omega '(d) - 2 \, l_f^2 \, {(\nabla \cdot \mathbf{A})}
			\cdot \nabla d + \mathbf{A}:\nabla(\nabla d)} \right]
		\label{30}
	\end{equation}
	
	\noindent Since $\mathbf{A}$ is a constant second-order structure tensor in this study. Therefore, the divergence of the structure tensor is derived as $\mathbf{\nabla \cdot A}=\bm{0}$. In the isotropic case, the structure tensor is equal to an identity tensor as $\mathbf{A}=\bm{I}$. Therefore, the phase-field fracture evolutionary equation is derived by substituting \hyperref[30]{Eq.\eqref{30}} into \hyperref[26]{Eq.\eqref{26}} and incorporating the two times gradient operator for the phase-field variable $d$ leading to $\mathbf{H} \, (d)$\footnote{The two times gradient operator towards $d$ leads to a Hessian matrix $\mathbf{H} \, (\bullet) = \nabla \, (\nabla (\bullet))$.} as:
	
	\begin{equation}
		\dot {d}=\frac{{d - {d_n}}}{{\Delta t}} = \frac{1}{{{\eta _f}}}{\left\langle {g'(d)\left[ {{\zeta} \left( {\frac{{\psi _e^ + }}{{{\psi _{th}}}} - 1} \right)} \right] + \frac{1}{{{c_0}}}\left[ {\omega '(d) - 2 \, l_f^2 \, \mathbf{H} \, (d) : {\mathbf{A}}} \right]} \right\rangle _ + } \quad {\rm{with}} \quad {\eta _f} = \frac{\zeta }{{2 \, {\psi _{th}}}}\eta. 
		\label{31}
	\end{equation}
	
	\noindent $d_n$ and $\Delta t$ in \hyperref[31]{Eq.\eqref{31}} denote the historical phase-field fracture variable and time increment, respectively. Here, the driving state function ${{\zeta} \left\langle {\frac{{\psi _e^ + }}{{{\psi _{th}}}} - 1} \right\rangle}_+$ in \hyperref[31]{Eq.\ref{31}} provides many possibilities to govern the fracture evolution, such as effective stress or energy-based criterion.
	\begin{itemize}
		\item Remark 3: Stress-based criterion
	\end{itemize}
	In this context, the effective stress tensor $\tilde{\bm{\sigma}}$ is employed with a definition of the effective operator $(\tilde{\bullet}) = (\bullet)/g(d)$. 
	\begin{equation}
		\tilde{\bm{\sigma}} =  \tilde{\bm{\sigma}}_+ + \tilde{\bm{\sigma}}_- , \quad \text{with} \quad \bm{\sigma} = \frac{\partial {\psi_{{\rm{bulk}}}}}{\partial \bm{\varepsilon}}   \quad \text{and} \quad {\bm{\sigma} _ \pm } := \sum\limits_{i = 1}^3 {{{\left\langle {{{\sigma} _i}} \right\rangle }_ \pm } \, {\bm{n}_i} \otimes {\bm{n}_i}}
		\label{31-1}
	\end{equation}
	Hence, the driving state function is expressed as ${{\zeta} \left\langle {\sum\limits_{i = 1}^3 \left(\frac{{{\left\langle {{{\sigma} _i}} \right\rangle }_ + }}{{{\sigma _{th}}}} \right)^2 - 1} \right\rangle}_+$ which represents the ratio between the principal stress ${{\left\langle {{{\sigma} _i}} \right\rangle }_ + }$ due to the tensile and the stress threshold $\sigma_{th}$.
	
	\begin{itemize}
		\item Remark 4: Energetic criterion with threshold
	\end{itemize}
	Here, the crack driving force functions for an energetic criterion are determined by \hyperref[8]{Eq.\eqref{8}}, which is based on Griffith’s critical energy release rate $G_c$ and describes the fracture density function as:
	\begin{equation}
		\dot d = \frac{{d - {d_n}}}{{\Delta t}} = \frac{1}{{{\eta _f}}}{\left\langle {\frac{{{l_f}}}{G_{c}}g'(d) \, {\mathcal{H}} + \frac{1}{{{c_0}}}\left[ {\omega '(d) - 2 \, l_f^2 \, \mathbf{H} \, (d) : {\mathbf{A}}} \right]} \right\rangle _ + } \quad {\rm{with}} \quad {\eta _f} = \frac{{{l_f}}}{G_{c}}\eta,
		\label{32}
	\end{equation}
	
	\noindent where $\eta \ge 0$ represents a material parameter characterizing the viscosity of fracture propagation. $\mathcal{H}$ denotes the historical crack driving force, ensuring irreversibility in the phase-field crack evolution by filtering out the maximum value between the positive elastic energy $\psi_{e}^{+}$ and the damage threshold $\psi_{th}$. In contrast, the conventional AT2 model suffers from damage initiation even under minimal loads at the initial loading state. However, in real material loading scenarios, cracks are not initiated under such tiny initial loading conditions.
	
	\begin{equation}
		\mathcal{H} = \mathop {\rm {Max}}\limits_{s \in [0, \, t]} \Big[\psi _e^ + (\bm{\varepsilon} ,t), \, {\psi _{th}} \Big] \ge 0 \quad {\rm{with}} \quad 
		{\psi _{th}} = \frac{1}{2}E \, \varepsilon _0^2 = \frac{{\sigma _0^2}}{{2 \, E}}
		\label{33}
	\end{equation}
	
	\noindent Hence, inspired by the continuum damage growth criterion by \citet{simo1987strain}, the damage threshold $\psi_{th}$ is employed to govern the energy 
	for the damage initiation at strain $\varepsilon _{0}$. Fracture initiation occurs only when $\psi_{e}^{+} \ge \psi_{th}$. The analytical relationship between various material parameters and fracture strength is expressed in \hyperref[34]{Eq.\eqref{34}} by \citet{tanne2018crack}.
	\begin{equation}
		{\sigma _0} = \frac{9}{{16}}\sqrt {\frac{E \, G_{c}}{{3 \, {l_f}}}}
		\label{34}
	\end{equation}
	
	\noindent In this study, the cohesive fracture strength $\sigma _0$ between the fiber and matrix is employed based on the homogenized cohesive fracture strength at the nanoscale, as shown in \hyperref[fig6]{Fig.7 (b)}. Furthermore, the historical crack driving force $\mathcal{H}_{n}$ must satisfy the Karush-Kuhn-Tucker (KKT) conditions, which are expressed as follows:
	
	\begin{equation}
		\psi _e^ +  - \dot {\mathcal{H}}_{n} \le 0, \, \dot {\mathcal{H}}_{n}  \ge 0, \, \dot {\mathcal{H}}_{n}  \, (\psi _e^ +  -\dot {\mathcal{H}}_{n} ) = 0.
		\label{35}
	\end{equation}
	
	\subsubsection{Finite element implementation}
	The implementation of the material model in this study starts from the pseudopotential density function. As illustrated in \hyperref[13]{Eq.\eqref{13}}, crack propagation is driven by the energy competition between bulk energy and crack surface energy. Therefore, the overall rate-type potential is formulated as the sum of internal energy and external energy:
	
	\begin{equation}
		{\mathit{\Pi} _e}\left( {\bm{\varepsilon} ,d,\bm{u}} \right) = \underbrace {\int_{\Omega_e}  {{\psi _{{\rm{bulk}}}}(\bm{\varepsilon} ,d) \, dV}  + \int_{\Omega_e}  {{\psi _{{\rm{frac}}}}(d,\nabla d; \phi ) \, dV} }_{{\mathit{\Pi} ^{e} _{\rm{internal}}}} - \underbrace {\int_{\Omega_e}  {\bar{\bm{f}} \cdot \bm{u} \, dV}  - \int_{\partial {\Omega _e}} {\bar{\bm{t}} \cdot \bm{u} \, d\mathit{\Gamma} } }_{{\mathit{\Pi} ^{e} _{\rm{external}}}},
		\label{36}
	\end{equation}
	
	\noindent where $\mathit{\Pi} _{\rm{internal}}^{e}$ and $\mathit{\Pi} _{\rm{external}}^{e}$ denote the internal and external energies of an element $e$. Once the energy of an element $\mathit{\Pi}_e$ is determined, the primary goal of the finite element method (FEM) is to find all suitable displacements $\bm{u}$ and phase-field fracture variable $d$ by minimizing ${\mathit{\Pi} _e}$. To achieve such a goal, the partial derivative of ${\mathit{\Pi} _e}$ is derived with respect to the nodal degrees of freedom $\bm{U}_{e} = {[\bm{u}, \, d]_e}$. This process is implemented using the finite element programming toolbox: \textit{AceGen} \cite{korelc2016automation}. Consequently, the element residual vectors and stiffness matrices can be formulated as follows:
	\begin{equation}
		\bm{R}_e = \frac{{\partial {\mathit{\Pi} _{e} \, (\bm{u},\, d)}}}{{\partial \bm{U}_e}} \quad {{\rm{and}}} \quad \bm{K}^e = \frac{{\partial \bm{R}_e \, (\bm{u}, \, d)}}{{\partial \bm{U}_e}}.
		\label{37}
	\end{equation}
	
	\noindent The detailed parts of the element stiffness matrices $\bm{K}^e$ are expressed as:
	\begin{equation}
		{\bm{K}}_{uu}^e = \frac{{{\partial ^2}{\mathit{\Pi} _e}}}{{\partial {{\bm{u}}^2}}}, \quad {\bm{K}}_{ud}^e = \frac{{{\partial ^2}{\mathit{\Pi} _e}}}{{\partial {\bm{u}} \partial {{{d}}} }}, 
		\quad {\bm{K}}_{du }^e = \frac{{{\partial ^2}{\mathit{\Pi} _e}}}{{\partial {{d}}\partial {\bm{u}}  }}, \quad {{K}}_{dd}^e = \frac{{{\partial ^2}{\mathit{\Pi} _e}}}{{\partial {{{{{d}}} }^2}}}.
		\label{eq37-1}
	\end{equation}
	
	\noindent By assembling all elements' residuals $\bm{R}_e$, stiffness matrices $\bm{K}^e$, and displacement increments $\Delta \bm{U}_e$, the global residual vector $\bm{R}$, global stiffness matrix $\bm{K}$, and displacement increment vector $\Delta \bm{U}$ are obtained as:
	
	\begin{equation}
		{
			{\bm{R} = \raise3pt
				\hbox{$\hbox{\scriptsize $N_e$}\atop{\hbox{\LARGEbsf A}\atop {\scriptstyle e=1}}$} {\bm{R}_e}}, \quad {\bm{K} = \raise3pt
				\hbox{$\hbox{\scriptsize $N_e$}\atop{\hbox{\LARGEbsf A}\atop {\scriptstyle e=1}}$} {\bm{K}^e}}, \quad {\bm{U} = \raise3pt
				\hbox{$\hbox{\scriptsize $N_e$}\atop{\hbox{\LARGEbsf A}\atop {\scriptstyle e=1}}$} {\bm{U}_e}}, \quad {\rm{and}} \quad {\bm{R} + \bm{K} \Delta \bm{U} = \bm{0}},
		}   
		\label{38}
	\end{equation}
	
	\noindent where operator $\hbox{$\hbox{\scriptsize $N_e$}\atop{\hbox{\LARGEbsf A}\atop {\scriptstyle e=1}}$} \left( \bullet \right)$ in \hyperref[38]{Eq.\eqref{38}} represents the assembly operator applied to all $N_e$ elements in a discretized domain. Then, the linearized numerical coupled system is formulated in matrix form as follows:
	
	\begin{equation}
		\underbrace{\left[ {\begin{array}{*{20}{c}}
					{{\bm{K}_{{uu}}}}&{{\bm{K}_{{ud}}}}\\
					{{\bm{K}_{{du}}}}&{{\bm{K}_{{dd}}}}
			\end{array}} \right]}_{\bm{K}} 
		\underbrace{\left\{ {\begin{array}{*{20}{c}}
					{\Delta \bm{u}}\\
					{\Delta \bm{d}}
			\end{array}} \right\}}_{\Delta \bm{U}} =  - \underbrace{\left\{ {\begin{array}{*{20}{c}}
					{{\bm{R}_{u}}}\\
					{{\bm{R}_{d}}}
			\end{array}} \right\}}_{\bm{R}}.
		\label{39}
	\end{equation}
	
	\noindent With the global residual vector $\bm{R}$ and stiffness matrix $\bm{K}$ at hand, a global Newton-Raphson algorithm with an adaptive pseudo-time solver is employed to solve the coupled system based on the global primary field $\bm{U}$ and its increment $\Delta \bm{U}$ in an iterative manner. This process incurs high computational costs. To make the simulations more efficient, a multi-field decomposed model order reduction technique 
	can be further employed to solve the nonlinear residual vector $\bm{R}$.
	
	\subsection{RVE homogenization}
	\label{RVE_homogenization}
	To bridge the micro and macro scales in a finite element approach, effective stress, and strain are homogenized to characterize the overall mechanical behavior of the RVE composite. In this study, focusing on the polymer below the glass transition temperature $T_g$, the stress at any point $\bm{x}$ inside our RVE is formulated as $\bm{\sigma} \, (\bm{x}, \bm{\varepsilon}) =  \bm{\mathbb {C}}: \bm{\varepsilon}$. $\bm{\mathbb{C}}$ denotes a fourth-order material tensor for composite. Based on the orthotropic material moduli, the components $C_{11}$, $C_{22}$, $C_{33}$, $C_{12}$, and $C_{21}$ of the RVE can be formulated as:
	
	\begin{equation}
		\left[ {\begin{array}{*{20}{c}}
				{{\sigma _{11}}}\\
				{{\sigma _{22}}}\\
				{{\sigma _{12}}}
		\end{array}} \right] = \left[ {\begin{array}{*{20}{c}}
				{{C_{11}}}&{C_{12}}&0\\
				{C_{21}}&{{C_{22}}}&0\\
				0&0&{C_{33}}
		\end{array}} \right]\left[ {\begin{array}{*{20}{c}}
				{{\varepsilon _{11}}}\\
				{{\varepsilon _{22}}}\\
				{2 \, {\varepsilon _{12}}}
		\end{array}} \right].
		\label{40}
	\end{equation}
	
	\noindent The relationship given in \hyperref[40]{Eq.\eqref{40}} can be expressed in terms of Young's moduli $E_{11}$ and $E_{22}$ in $X$ and $Y$ directions. The Poisson's ratios $\nu_ {12}$ and $\nu_ {21}$ are determined through uniaxial tensile tests on specimens with a prescribed displacement of $u_{1}=0.001$. $G_{12}$ represents the shear modulus for the RVE under pure shear loading along the corresponding directions with a prescribed loading displacement ${u_1}/\sqrt 2$.
	
	\begin{equation}
		\left[ {\begin{array}{*{20}{c}}
				{{\sigma _{11}}}\\
				{{\sigma _{22}}}\\
				{{\sigma _{12}}}
		\end{array}} \right] = \underbrace{\frac{1}{{1 - {\nu _{12}} \, {\nu _{21}}}}\left[ {\begin{array}{*{20}{c}}
					{{E_{11}}}&{{E_{11}} \, {\nu _{21}}}&0\\
					{{E_{22}} \,{\nu _{12}}}&{{E_{22}}}&0\\
					0&0&{{G_{12}} \, (1 - {\nu _{12}} \, {\nu _{21}})}
			\end{array}} \right]}_{\mathbb{C}_{\text{RVE, vogit}}} \left[ {\begin{array}{*{20}{c}}
				{{\varepsilon _{11}}}\\
				{{\varepsilon _{22}}}\\
				{2 \, {\varepsilon _{12}}}
		\end{array}} \right]
		\label{41}
	\end{equation}
	
	\noindent For a two-phase composite, the homogenized stress $\left\langle \bm{\sigma}  \right\rangle$ and strain $\left\langle \bm{\varepsilon}  \right\rangle$ of the RVE are computed by integrating the stresses of each element and averaging over the volume of the entire RVE domain. The detailed equations are expressed in \hyperref[42]{Eq.\eqref{42}}.
	
	\begin{equation}
		{\left\langle \bm{\sigma}  \right\rangle}  = \frac{1}{V_{\rm{RVE}}} \int_\Omega  \bm{{\tilde \sigma }} \, dV  = \frac{1}{V_{\rm{RVE}}} \sum\limits_{e = 1}^{N_{e}} {\bm{{\tilde \sigma }} \, {V}}, \quad \left\langle \bm{\varepsilon}  \right\rangle  = \frac{1}{V_{\rm{RVE}}} \int_\Omega  {\bm{\tilde {\varepsilon }} \, dV}  = \frac{1}{{{V_{{\rm{RVE}}}}}}\sum\limits_{e = 1}^{N_{e}} {\bm{\tilde \varepsilon} \, {V}} 
		\label{42}
	\end{equation}
	
	\noindent where ${\bm{{\tilde \sigma }}}$ and ${\bm{{\tilde \varepsilon }}}$ represent the effective stress and strain of one element. $V_{\rm{RVE}}$ and $ V$ denote the volume of RVE and the volume of an element, respectively. $\bm{\Bbb {C}}_{{\rm{RVE,vogit}}}$ represents the Voigt bound on the RVE elastic moduli of the effective homogenized RVE composite. This bound establishes an upper response on the linear elastic behavior of the RVE composite, representing a single Gaussian point at the macroscale.
	In \hyperref[41]{Eq.\eqref{41}}, $E_{11}={\left\langle \bm{\sigma}_{11}  \right\rangle} /{\left\langle \bm{\varepsilon}_{11}  \right\rangle}$, $E_{22}={\left\langle \bm{\sigma}_{22}  \right\rangle} /{\left\langle \bm{\varepsilon}_{22}  \right\rangle}$, $\nu_{12}=-{\left\langle \bm{\varepsilon}_{22}  \right\rangle} /{\left\langle \bm{\varepsilon}_{11}  \right\rangle}$, and $\nu_{21}=-{\left\langle \bm{\varepsilon}_{11}  \right\rangle} /{\left\langle \bm{\varepsilon}_{22}  \right\rangle}$.
	
	\section{Cohesive phase-field model validation}
	\label{Sec3: Model validation}
	\subsection{Isotropic three-point bending test}
\begin{figure*}[!h]
		\centering
		\includegraphics[width=18cm]{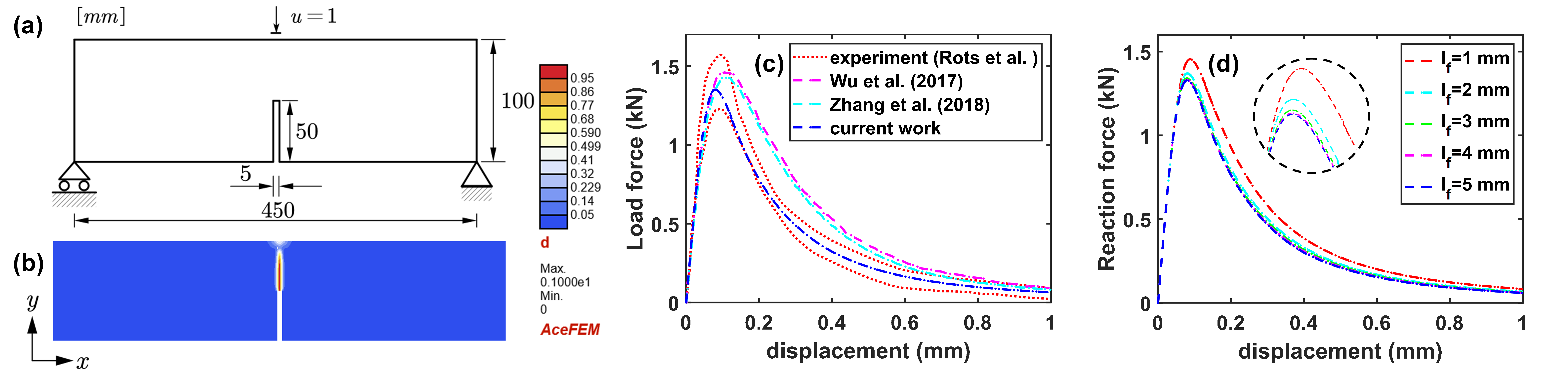}
		\caption{(a) Geometry and boundary conditions for the three-point bending test using an isotropic cohesive phase-field fracture model, with a specimen thickness of $100$ mm. (b) Contour plot showing phase-field fracture results with a length scale parameter $l_{f} = 2.5 \, \text{mm}$. (c) Comparison of force-displacement curves with reference studies. (d) Study of the length scale parameter $l_{f}$, varying from 1 mm to 5 mm.}
		\label{fig2}
	\end{figure*}

	To validate the current material model in a physical manner, this study compares the force-displacement curves (see \hyperref[fig2]{Fig.3 (c)}) with experimental data from \citet{rots1988computational}, the non-tension/compression asymmetric decomposed cohesive phase-field fracture model proposed by \citet{wu2017unified}, and the non-dissipated cohesive model proposed by \citet{zhang2018iteration}. The selected parameters are as follows: elastic modulus $ E = 2.0 \times 10^{4} \, \text{MPa} $, Poisson's ratio $ \nu = 0.2 $, critical energy release rate $ G_{c} = 0.113 \, \text{N/mm} $, phase-field regularized length scale $ l_{f} = 2.5 \, \text{mm} $, cohesive fracture strength $ \sigma_{0} = 2.4 \, \text{MPa} $, the star-convex decomposition controlling parameter $\gamma^* = 3$, and the \citet{cornelissen1986experimental} softening parameters which characterize the cohesive softening response ($ p = 2 $, $ a_{2} = 1.3868 $, $ a_{3} = 0.6567 $, see \hyperref[18]{Eq.\eqref{18}} and \cite{wu2017unified}), and the viscosity of crack propagation $ \eta_{f} = 1.0 \times 10^{-6} \, \text{Ns/mm}^2 $. To emphasize the preference for the cohesive regularized phase-field model over the standard phase-field fracture models (AT2), which suffer from issues related to the diffused fracture length scale dependency, a length scale convergence study is conducted with $l_f$ varied from 1 mm to 5 mm (see \hyperref[fig2]{Fig.3 (d)}).
	
	The results in \hyperref[fig2]{Fig.3 (c)} confirm that the current material model not only demonstrates improved numerical performance compared to the non-dissipative models \cite{rezaei2022anisotropic,nguyen2018modeling} but also aligns well with experimental fracture softening data \cite{rots1988computational}. This improvement is attributed to the incorporation of asymmetric tension-compression decomposition in the strain energy density function, as specified in \hyperref[19]{Eq.\eqref{19}} and \hyperref[20]{Eq.\eqref{20}}, and the viscosity-enhanced fracture density function in \hyperref[32]{Eq.\eqref{32}}. Subsequently, the anisotropic numerical benchmark tests are discussed in \hyperref[S3-2]{Section 3.2}.
	
	\subsection{Anisotropic three-point bending test}
	\label{S3-2}
\begin{figure*}[!h]
		\centering
		\includegraphics[width=18cm]{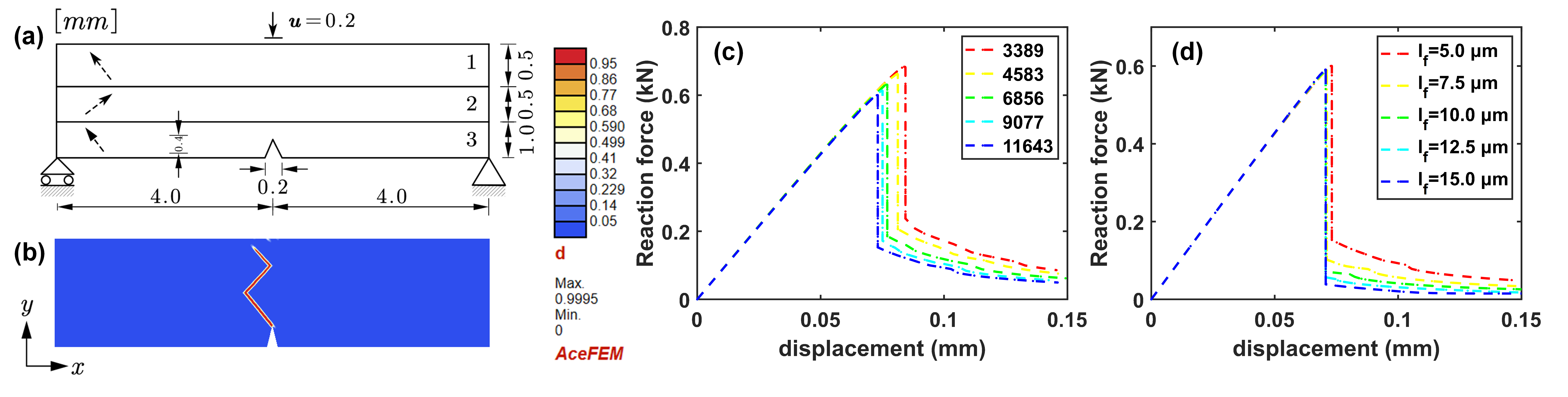}
		\caption{(a) Geometry and boundary value problem of the anisotropic three-point bending test with a predefined triangular notch. The dashed arrows represent the predefined anisotropic angles $\phi_{1} = {3\pi}/{4}$ and $\phi_{2} = {\pi}/{4}$; (b) contour plot of the phase-field fracture trajectory in the anisotropic bending test conducted by \textit{AceFEM} \cite{korelc2016automation}; (c) mesh convergence tests; (d) comparison of the lengths for the diffused damage band from 5 $\mu \rm m$ to 15 $\mu \rm m$.}
		\label{fig3}
	\end{figure*}

	For the anisotropic bending test (see \hyperref[fig3]{Fig.4 (a)}), the material parameters are defined as: elastic modulus $E = 2.1 \times 10^5 \, \text{MPa}$, Poisson’s ratio $\nu = 0.3 $, length scale $l_{f} = 5$ $\mu$m, the viscosity of the crack propagation $\eta_{f} = 1.0 \times 10^{-6} \, \text{Ns/mm}^2$, critical energy release rate $G_{c} = 2.7 \, \text{N/mm}$, the star-convex decomposition parameter $\gamma^* = 3$, the first predefined anisotropic fracture angle $\phi_{1} = {3\pi}/{4}$, the second predefined anisotropic fracture angle $\phi_{2} = {\pi}/{4}$, and penalty parameter $\alpha = 500$. Additionally, the linear softening parameters characterizing the cohesive response are $p = 2$, $a_{2} = -{1}/{2}$, and $a_{3} = 0$ (see \hyperref[18]{Eq.\eqref{18}}). The analytical fracture strength $\sigma_{0}$ is evaluated using \hyperref[34]{Eq.\eqref{34}}. The fracture trajectory of the bending test closely aligns with both the classical AT2 model proposed by \citet{miehe2010thermodynamically} and the anisotropic fracture model investigated by \citet{teichtmeister2017phase}. It should be noted that the choice of the penalty parameter remains a topic of debate within the phase-field community. \citet{zhang2019phase} have proposed defining this parameter as a model penalty parameter (MPP) based on the ratio between the strong $G_s$ and the weak $G_w$ energy release rates for fiber-reinforced laminate composites as follows:
	
	\begin{equation}
		\alpha = (\frac{G_s}{G_w})^{2}-1.
		\label{43}
	\end{equation}
	
	\noindent Using \hyperref[43]{Eq.\eqref{43}}, the penalized parameter $\alpha = 248$ ($G_{s} = G_{c}^{\text{GN}}, \, G_{w} = G_{c}^{\text{P3HT}}$) is calculated for the anisotropic bending test (\hyperref[fig3]{Fig.4}). The results indicate that this value of $\alpha$ does not sufficiently penalize the predefined anisotropic angles $3\pi /4$ and $\pi /4$. Therefore, in this study, $\alpha$ is considered to be a flexible numerical parameter. By conducting anisotropic three-point bending tests (see \hyperref[fig3]{Fig.4 (a)}), it is observed in \hyperref[fig3]{Fig.4 (b)} that the current model aligns well with several reference studies conducted by \citet{teichtmeister2017phase}. Additionally, studies on mesh convergence and phase-field diffused fracture band lengths are shown in \hyperref[fig3]{Fig.4 (c)} and \hyperref[fig3]{Fig.4 (d)}, ranging from 5 $\mu \rm m$ to 15 $\mu \rm m$, respectively. The results in \hyperref[fig3]{Fig.4 (d)} suggest that the fracture length parameter converges as $l_f$ reaches a specific value in anisotropic bending fracture tests. Thus, the fracture behavior of GN and $\text{C}_{x}\text{N}_{y}$ reinforced RVE composites is investigated in the following sections.
	
	\newpage
	\section{Results and discussion}
	\label{Sec4: Results and discussion}
	\subsection{Nanoscale: mechanical/fracture study of fibers, matrix, and their cohesive interactions}
	\label{nanoscale}
	
	\begin{figure*}[!h]
		\centering
		\includegraphics[width=18cm]{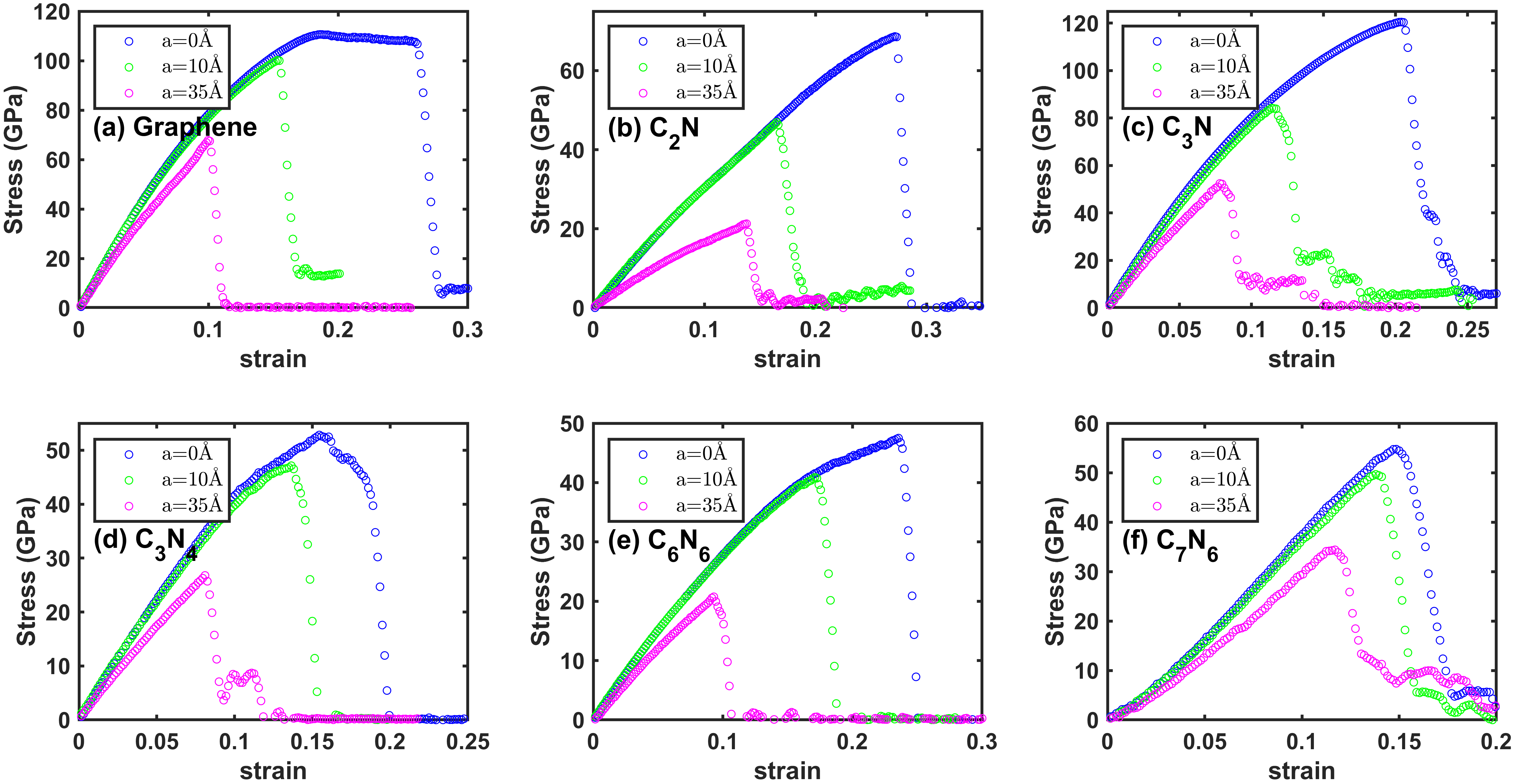}
		\caption{The homogenized stress-strain response of GN and $\text{C}_{x}\text{N}_{y}$ monolayers for uniaxial tensile tests along the armchair direction at 300 K with predefined notches of 0 \text{\r{A}}, 10 \text{\r{A}}, and 35 \text{\r{A}}, where the predefined notches are located in the zigzag direction. Here, $a = 0$ \text{\r{A}} indicates the pristine monolayers without a predefined notch.}
		\label{fig4}
	\end{figure*}
	
	A molecular dynamics simulation of a rectangular supercell with three predefined notches is presented herein. A non-notched ($a = 0$ \text{\r{A}}) specimen (\hyperref[fig1]{Fig.1 (a)}) is employed to calculate the elastic modulus $E_{\text{fiber}}$ and Poisson's ratio $\nu_{\text{fiber}}$ of the fibers (GN and $\text{C}_{x}\text{N}_{y}$) at a low strain level of 0.01. Most of the evaluated elastic moduli (GN, $\text{C}_2\text{N}$, $\text{C}_3\text{N}$, and $\text{C}_7\text{N}_6$) demonstrate good agreement with the reference works, with the exception of the Poisson's ratio (see \hyperref[Table-1]{Table 1}). The variation in Poisson's ratio can be attributed to the out-of-plane deflection and wrinkling induced by three-body atomic fluctuations from $\varsigma_{ij}^{m} (\theta_{ijk})$ in the \textit{Tersoff} potential (see \hyperref[AppendixA]{Appendix A}). However, Poisson's effect exerts a minimal influence on the microscale mechanical response, as the phase-field fracture model operates under the assumption of a small strain. Furthermore, the calculated critical energy release rates demonstrate good agreement with the studies in \cite{zhang2014atomistic,shishir2019molecular} based on the stress-strain curves in \hyperref[fig4]{Fig.5} and \hyperref[2]{Eq.\eqref{2}}. For brevity, this investigation focuses exclusively on the mechanical properties along the armchair direction of pristine GN and $\text{C}_x\text{N}_y$ monolayers. As anticipated from \hyperref[fig4]{Fig.5}, the ultimate tensile strength (UTS) decreases as the length of the predefined notches increases. The calculated elastic moduli of the fibers from GN to $\text{C}_x\text{N}_y$ are 999.56, 996.06, 347.85, 487.19, 326.25, and 662.50 GPa. The rounded parameters are presented as follows.
	
	\begin{table}[!h]
		\begin{tabular}{lcccl}
			\hline
			\multicolumn{1}{c}{}  & $E$ (GPa) & $\nu$ & $G_c$ ($J/m^2$) & \multicolumn{1}{c}{Reference works and experiments}                                 \\ \hline
			P3HT                 & 1.24                 & 0.40           & 2.10                        & $\nu=0.3$ \cite{menichetti2017strain}, 0.35 \cite{tahk2009elastic}, $G_{c}=2.10$  \cite{kim2022numerical}                                      \\ \hline
			GN                   & 1000                & 0.11           & 33.09                     & $\nu=0.147$ \cite{gupta2005elastic}, 0.16 \cite{awasthi2008modeling}, 0.173 \cite{gui2008band}, $G_{c}=33.02$ \cite{zhang2014atomistic}, $E=1.0 \pm 0.1$ TPa \cite{lee2008measurement} \\
			$\rm{C_{3}N}$                  & 996                & 0.13           & 12.07                     & $\nu=0.155$ \cite{zhou2017computational}, $E=978.75 \pm 4.7$ GPa 
			, $G_{c}=10.16$ \cite{shishir2019molecular}, 10.98 \cite{shishir2021investigation}      \\
			$\rm{C_{2}N}$                   & 348                & 0.36           & 15.80                     & $E=343.34$ GPa \cite{abdullahi2018elastic}, $335 \pm 5$ GPa 
			                                          \\
			$\rm{C_{3}N_{4}}$                  & 487                & 0.19           & 18.52                     &                                                                                     \\
			$\rm{C_{6}N_{6}}$                  & 326                & 0.23           & 18.64                     &                                                                                     \\
			$\rm{C_{7}N_{6}}$                  & 663                & 0.42           & 8.95                      & 
			\\ \hline
		\end{tabular}
		\caption{Homogenized elastic modulus $(E)$, Poisson's ratio $(\nu)$, the energy release rate $(G_{c})$ of fibers in the armchair direction as well as the P3HT matrix.}
		\label{Table-1}
	\end{table}
	
\begin{figure*}[!h]
		\centering
		\includegraphics[width=18cm]{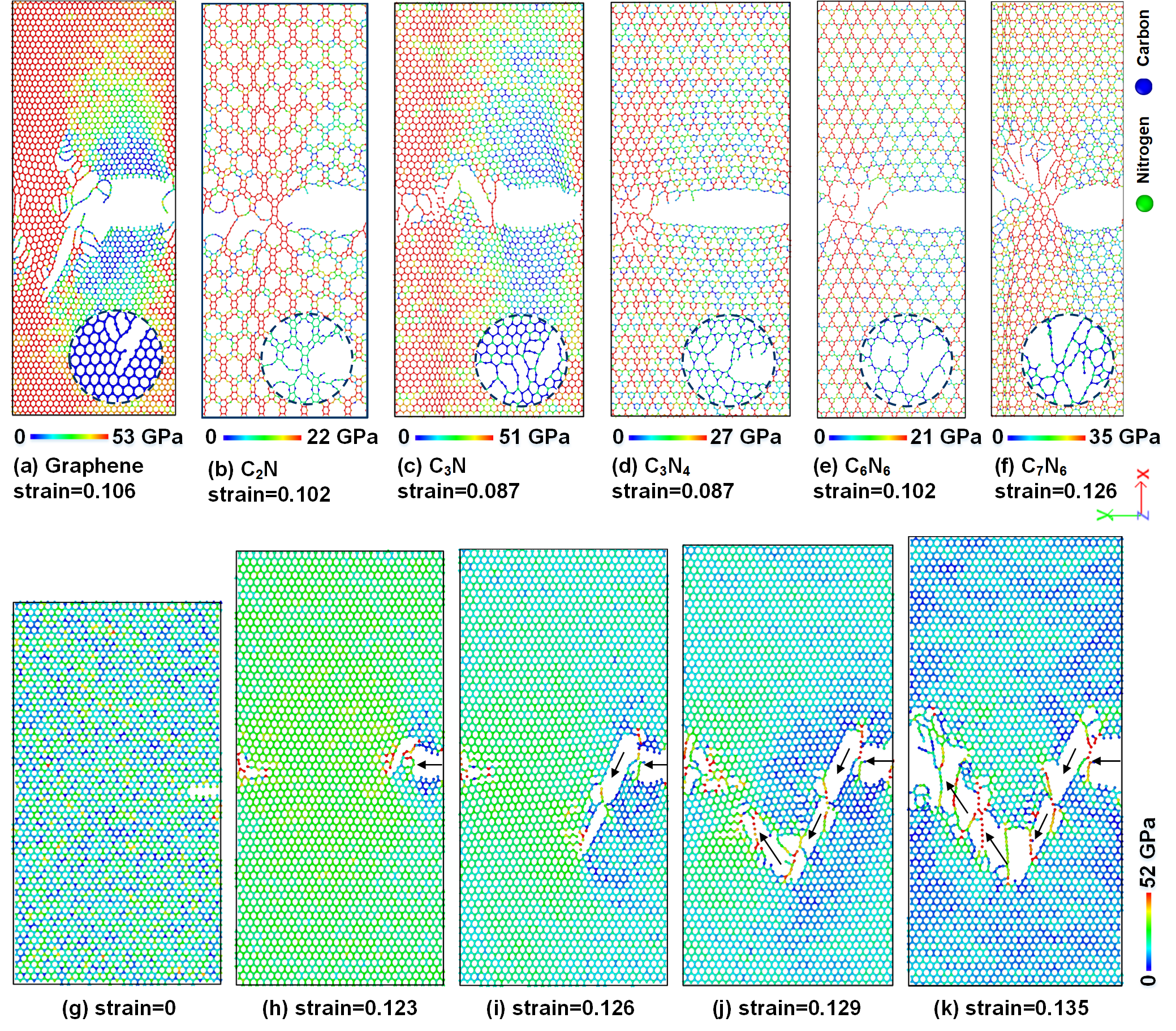}
		\caption{(a-f) Schematic illustration of normal stress $({\sigma}_{xx})$ distribution; the crack tips are enlarged for GN and $ \text{C}_{x}\text{N}_{y} $ nanosheets with an initial crack length of $35$ \text{\r {A}}; (g-k) von Mises stress contour plots and the anisotropic crack evolution of a $\rm{C_{3}N}$ nanosheet under uniaxial tensile testing with a predefined crack length of $10$ \text{\r {A}}. The arrow inside the graph indicates the direction of the crack evolution.}
		\label{fig5}
	\end{figure*}

	\newpage
	The observations gleaned from \hyperref[fig4]{Fig.5} and \hyperref[fig5]{Fig.6} are of significant interest, revealing a notable trend: as the atomic percentage of nitrogen increases from $33.33\%$ in $ \text{C}_2\text{N} $ to $46.16\%$ in $\text{C}_7\text{N}_6$, the ultimate fracture strength of the monolayers progressively decreases. This phenomenon can be attributed to the hypothesis that the increase of C-N bonds enhances the ductile fracture characteristics of the monolayer, particularly at higher strain levels (see \hyperref[fig5]{Fig.6}). When $\text{C}_{x}\text{N}_{y}$ monolayers are subjected to tension near the critical fracture point, the increased number of more ductile C-N bonds (owing to the increased nitrogen content) contributes to a higher level of fracture strain to reach the fracture point of the carbon-nitride monolayers. Considering this factor, a substantial increase in the length of the stretched specimen in physical phenomena is observed, even when the material approaches its critical fracture strength, as illustrated in \hyperref[fig5]{Fig.6 (c)} and \hyperref[fig5]{Fig.6 (f)}.
	
	\FloatBarrier
	
	Furthermore, the intrinsic honeycomb lattice structure of the nanolayer material facilitates the formation of anisotropic patterns in the evolution of cracks. It is evident that the aforementioned fracture trajectories are influenced not only by the relatively weaker C-N bond potentials but also by the structural weakening mechanisms that arise from the honeycomb lattice structure and vacuum effects at the crack front. Notably, the crack trajectories of GN and $\text{C}_3\text{N}$ exhibit a distinct "$\mathbf{{V}}$" shape in \hyperref[fig5]{Fig.6 (g-k)}, demonstrating pronounced anisotropic fracture characteristics. These specific anisotropic behaviors of GN and $\text{C}_3\text{N}$ were observed experimentally, as reported by \citet{qu2022anisotropic}.
	
	Upon detailed analysis of the fracture region of the monolayer in \hyperref[fig5]{Fig.6 (g-k)}, it is observed that the crack propagates continuously around the boundary edge. Initially, cracks initiate from the notch tip and evolve with a crack orientation of $\pi/3$ owing to the honeycomb lattice structure of $\text{C}_3\text{N}$ (\hyperref[fig5]{Fig.6 (c)}). Simultaneously, a small crack also initiates around the left-hand side, see \hyperref[fig5]{Fig.6 (h)}. This phenomenon arises because of the periodic boundary conditions imposed in both the $X$ and $Y$ directions. Consequently, the crack growth trajectory aligns with the pattern observed in studies proposed by \citet{rezaei2019atomistically} and \citet{cao2020novel}, which stands in contrast to the discontinuous fracture trajectory in both $X$ and $Y$ directions.
	
	\FloatBarrier
	\begin{figure*}[!h]
		\centering
		\includegraphics[width=18cm]{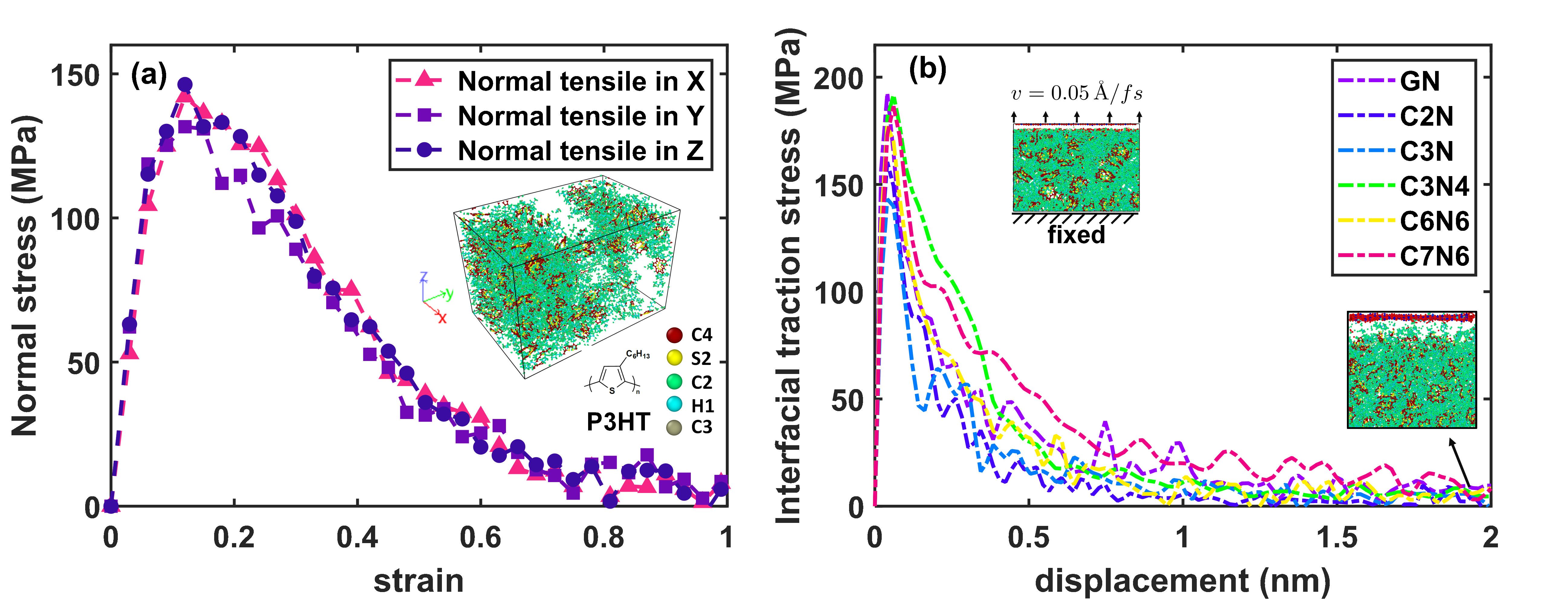}
		\caption{(a) Schematic illustration of the stress-strain response for P3HT matrix under uni-axial tensile testing along $X$, $Y$, and $Z$ directions at 300K. The cubic specimen is designed as an RVE with a length of 60.1 \text{\AA}. The atomic structure illustrates the fracture of the P3HT polymer cube during the test, with C4 (naphthenic-carbon), C2, C3 (carbon), S2 (sulfur), and hydrogen (H1) denoting the corresponding atoms of P3HT. (b) Investigation of interfacial cohesive strength between fiber and matrix through normal traction of the single layer.}
		\label{fig6}
	\end{figure*}
	
	To investigate the mechanical performance of the P3HT matrix, uniaxial tensile tests are conducted along the three axes ($X$, $Y$, and $Z$). The stress-strain relationship in \hyperref[fig6]{Fig.7 (a)} indicates that P3HT exhibits isotropic characteristics in both the elastic and softening regions. The cohesive interaction between the fiber and the matrix is further investigated through traction-separation ($\sigma _{\rm{coh}}-w$) relationships, as shown in \hyperref[fig6]{Fig.7 (b)}. Based on the traction-separation data in \hyperref[fig6]{Fig.7 (b)}, the homogenized fracture strengths of GN to $\rm{C_{7}N_{6}}$ are calculated as 193, 151, 141, 190, 177, and 186 MPa, respectively.
	
	It is observed in \hyperref[fig6]{Fig.7 (b)} that as soon as the traction displacement reaches $1.25 \, \text{nm}$ (12.5 \text{\AA}), the homogenized cohesive strength decreases nearly to zero, consistent with the predefined cutoff $r_{c} = 12.0 \, \text{\AA}$ in \hyperref[5]{Eq.\eqref{5}}. The cohesive response of GN/P3HT shares a similar traction-separation behavior observed in GN/polyethylene by \citet{awasthi2008modeling}, as well as the analytical cohesive solution proposed by \citet{lu2008cohesive}. Interestingly, there is a significant difference (approximately 60 MPa) between the fracture strength of the pure matrix and the cohesive fracture strength. This indicates that the \textit{van der Waals}-induced interfacial interactions play a crucial role in enhancing the mechanical properties of GN and $\text{C}_{x}\text{N}_{y}$ reinforced nanocomposites, as observed in our previous studies 
	. However, this cohesive-induced enhancement is validated only within the predefined cutoff between the fiber and matrix at the interface, see \hyperref[AppendixB]{Appendix B}. Beyond the predefined cutoff, the cohesive energy-induced traction force is negligible.
	
	\subsection{$C_{x}N_{y}/P3HT$ reinforced composite fracture}
	In this section, the homogenized material data from \hyperref[Table-1]{Table 1} are utilized to investigate the mechanical and fracture behaviors of the fiber-reinforced composites at the microscale. 
	It is noteworthy that Poisson's ratios from established reference works are employed when substantial discrepancies arise between the calculated results and reference data. Square RVEs with fiber volume fractions of $1\%$, $2\%$, and $3\%$ are designed to simulate reinforced composites with rectangular inclusions, representing the cross-sectional structure of three-dimensional plate fiber-reinforced composites. The uniaxial tensile tests are conducted to calculate the effective modulus and ultimate tensile strength (UTS) of ${\rm{GN}}/\text{C}_{x}\text{N}_{y}$ reinforced P3HT composites. The fibers are modeled with anisotropic behavior, oriented at a predefined angle $\phi = \pi/3$, see \hyperref[fig5]{Fig.6}. The penalty parameter and phase-field fracture length scale are set as $\alpha = 500$ and $l_f = 0.01$, respectively. The tension and compression-based star-convex energy decomposition parameter is defined as $\gamma^* = 3$ to control the ratio between the compression and tension terms in the stress. Mechanical parameters (elastic modulus $E$, Poisson's ratio $\nu$, and critical energy release rate $G_c$) for both the fibers and matrix are employed from \hyperref[Table-1]{Table 1}. Additionally, the cohesive fracture strength $\sigma_{0}$, which governs the cohesive behavior between the fiber and matrix, is obtained using a hybrid potential in molecular dynamics simulations. A linear cohesive softening performance is employed with the parameters $p = 2$, $a_2 = -1/2$, and $a_3 = 0$, as described in \hyperref[18]{Eq.\eqref{18}}. 
	
	\begin{figure}[!h]
		\centering
		\includegraphics[width=18cm]{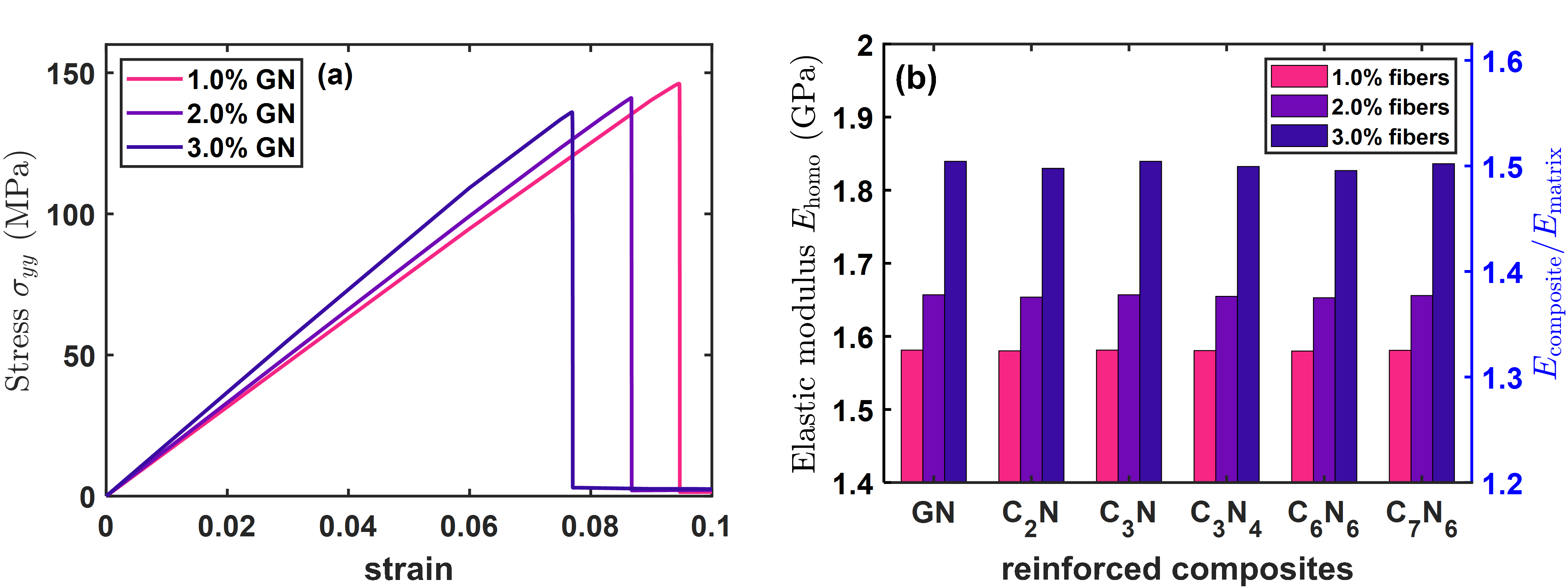}
		\caption{(a) Force-displacement curves comparison in the case of $1\%$, $2\%$, and $3\%$ graphene (GN) reinforce composites. (b) Homogenized elastic modulus and enhancement ($E_{\text{composite}}/E_{\text{matrix}}$) comparison between graphene reinforced composites and $\text{C}_{x}\text{N}_{y}$ reinforced composites.}
		\label{fig7}
	\end{figure}
	
	As shown in \hyperref[fig7]{Fig.8 (a)}, as the volume fraction of graphene increases from $1\%$ to $3\%$, the critical fracture strain decreases gradually. This indicates that the composites become stiffer and more brittle, which is attributed to enhancements from both the graphene itself and its interfacial contribution. \hyperref[fig7]{Fig.8 (b)} illustrates that the homogenized elastic modulus increases consistently with an increase in the fiber's volume fraction. Among the various reinforcements, graphene exhibits superior performance in enhancing the homogenized elastic modulus compared with other fibers. Notably, the $\rm{C_3N}$ reinforced composites show a similar enhancement to graphene-reinforced composites, reaching 1.486 times stiffer than the pure matrix when the fiber's volume fraction reaches $3.0\%$. This enhancement can be attributed to the analogous mechanical performance of the graphene and $\rm{C_3N}$ monolayers, as depicted in \hyperref[fig4]{Fig.5 (a)} and \hyperref[fig4]{Fig.5 (c)}. The homogenized elastic modulus of $3\%$ graphene and carbon-nitride ($\text{C}_{x}\text{N}_{y}$) reinforced P3HT composites range between 1.827 and 1.839 GPa.
	
	\begin{figure*}[!h]
		\centering
		\includegraphics[width=18cm]{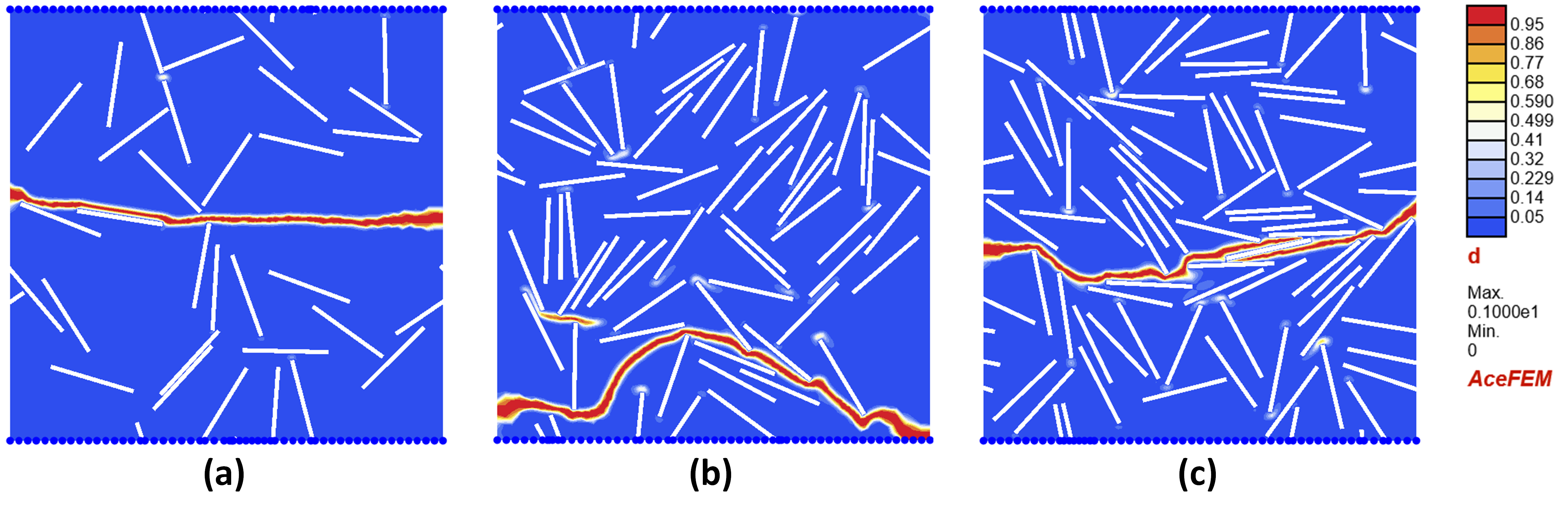}
		\caption{Phase-field fracture of graphene-reinforced P3HT composites for fiber's volume fraction of (a) $1\%$, (b) $2\%$, and (c) $3\%$ at the microscale. The white area represents the fibers and the rest part denotes the matrix.}
		\label{fig8}
	\end{figure*}
	
	\hyperref[fig8]{Fig.9} shows the phase-field fracture results of the graphene-reinforced RVE composites with various volume fractions ($1\%$, $2\%$, and $3\%$). Fracture trajectories predominantly occur within the matrix, particularly around the interface region between the fiber and the matrix. This behavior is attributed to cohesive interactions between the fiber and matrix, governed by \textit{van der Waals} potentials. (\hyperref[5]{Eq.\eqref{5}}), as shown in \hyperref[fig6]{Fig.7}. 
	
	While a dominant crack trajectory evolves throughout the entire RVE, there are also some initiations for the secondary damage trajectories within the specimen, as observed in \hyperref[fig8]{Fig.9 (b)}. Here, the secondary damage trajectories are regarded as cracks that possess the phase-field variable ($d$) between 0.5 and 0.8. These non-dominant cracks primarily occur around the edge tip of the fiber, where strong discontinuities are present in the composite. These discontinuities lead to large stress concentrations in these regions even though a tiny external load is applied to the specimen. Consequently, crack initiation is promoted due to these stress concentrations at the edge tip of the fiber. This also demonstrates that when the stress-governed damage evolution \hyperref[31-1]{Eq.\eqref{31-1}} is employed, the unavoidable stress concentration will dominate crack evolution with a tiny loading. In other words, this also explains why the energy-governed crack evolution in \hyperref[32]{Eq.\eqref{32}} is preferred for composite fracture modeling in this study. This is because the predefined damage threshold $\psi_{th}$ will avoid the nonphysical damage initialization at a tiny external loading. This somehow demonstrates that, when designing composite materials, a smooth geometrical inclusion inside the composite will enhance the mechanical properties by avoiding geometrical discontinuity-induced stress concentration. Indeed, these discontinuities can facilitate the initiation of randomly distributed secondary damage trajectories within the specimen, thereby leading to degradation in the mechanical performance of the composite.
	
	\subsection{Homogenization and scaling-effect}
	Following an in-depth analysis of fracture behavior in RVEs, investigating both cohesive and non-cohesive factors is essential to deeply understand the mechanical and fracture responses of graphene and carbon nitride-reinforced composites. Key factors such as fiber/matrix interfacial strength and fiber loading orientation significantly influence the fracture strength of micro- and nano-fiber-reinforced composites. Accordingly, a comparative study of the homogenized UTS between cohesive and non-cohesive (standard AT2) models is presented in the following section.
	
\begin{figure*}[!h]
		\centering
		\includegraphics[width=18cm]{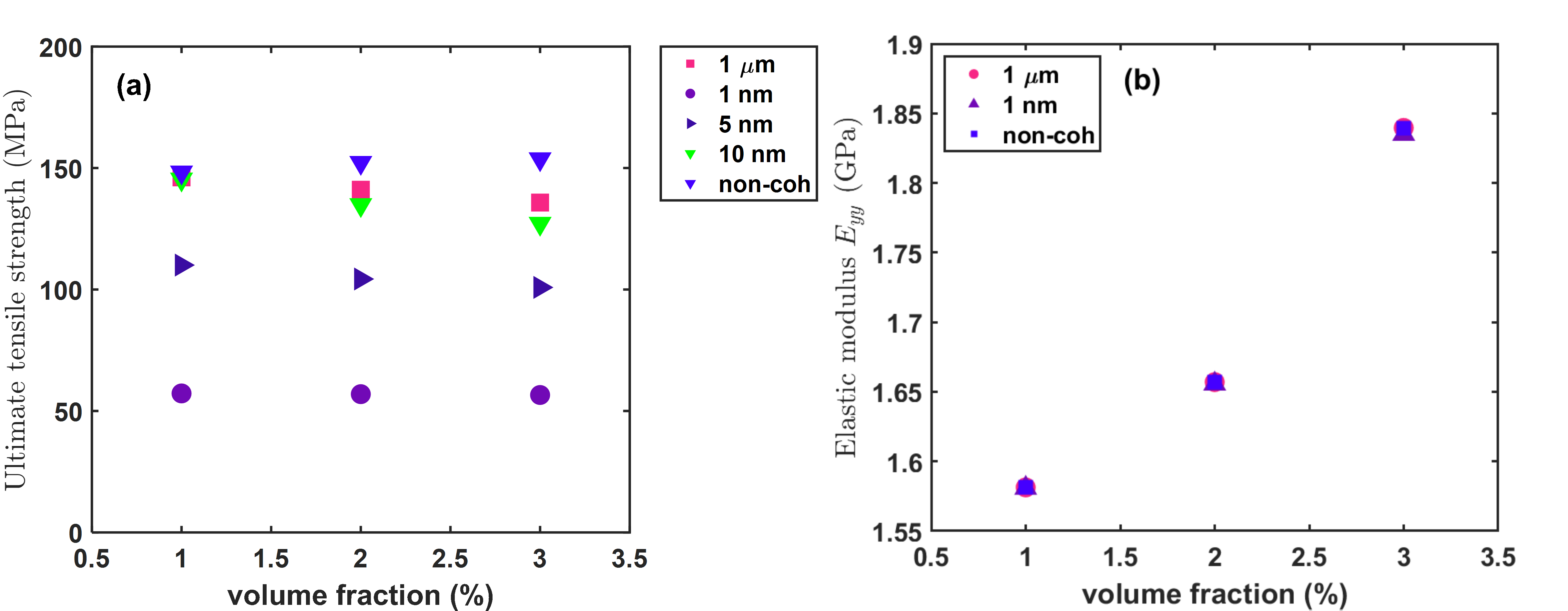}
		\caption{(a) Fracture strengths comparison for scale size from $\rm{1\,nm}$ to $\rm{1\,\mu m}$ as well as non-cohesive model (AT2, $\rm{1\,\mu m}$) for volume fraction range from $1\%$ to $3\%$. (b) Homogenized elastic modulus comparison for scale size of $\rm{1\,nm}$ to $\rm{1\,\mu m}$, and non-cohesive model.}
		\label{fig9}
	\end{figure*}

	As illustrated in \hyperref[fig9]{Fig.10}, the UTS in the non-cohesive model consistently increases as the fiber volume fraction increases. In contrast, the cohesive model demonstrates a different trend, with the UTS gradually decreasing as the fiber volume fraction increases. This discrepancy arises from the growing influence of interfacial interactions with the expansion of the fiber interface, which increasingly affects the fracture mechanism of the composite. Additionally, the non-cohesive model may overestimate the UTS, 
	leading to a reduced disparity between the cohesive and non-cohesive models at lower fiber volume fractions. In composites with fewer inclusions, the interfacial interaction between the fiber and matrix becomes negligible, resulting in a smaller contribution from the \textit{van der Waals}-induced cohesive energy relative to the reinforcement effect provided by the fibers. Thus, for composites with lower reinforcement volume fractions, the UTS of the RVE remains similar in both models. This trend is observed not only in the present study (see in \hyperref[fig9]{Fig.10 (a)} for $1\%$ graphene), but also in previous work 
	, which combined a continuum damage model with a displacement-jump-based cohesive zone model. Furthermore, the experimental data reported by \citet{fu2008effects} also confirm the response of the UTS across various fiber volume fractions.
	
	Another critical phenomenon in multiscale modeling is the scaling effect, which is observed in simulations of thermal conductivity 
	and mechanical properties 
	, as well as in experiments conducted by \citet{malekpour2014thermal}. As illustrated in \hyperref[fig9]{Fig.10 (a)}, the scaling effect is significant in both the homogenized UTS and elastic modulus for the cohesive fracture model, spanning from the nanoscale to the microscale. When comparing scales of 1 nm, 5 nm, 10 nm, and 1 $\rm{\mu m}$ at a consistent volume fraction, the UTS of the graphene-reinforced composites decreases as the scale size decreases from 1 $\mu$m to 1 nm. Interestingly, \hyperref[fig9]{Fig.10 (a)} shows that the primary scaling effect on the UTS occurs between 1 nm and 10 nm, with mean differences of $45.78\%$ between 1 nm and 5 nm as well as $22.54\%$ between 5 nm and 10 nm. This substantial variation across scales underscores the importance of carefully considering the scaling effect when modeling and designing composite materials from small to large scales. Thus, it is essential to recognize that the performance of composite materials is influenced not only by the complex physical properties of the fibers, matrix, and interface but also by the scaling effect on the UTS across nanoscale to microscale dimensions. 
	
	In addition to the UTS, the homogenized elastic modulus is essential for characterizing the mechanical behavior of composites. As depicted in \hyperref[fig9]{Fig.10 (b)}, the scaling effect on the homogenized elastic modulus is minimal, with the elastic modulus at the nanoscale only marginally lower than at the microscale. This difference can typically be disregarded when the fiber volume fraction is relatively low (below $2\%$). Furthermore, comparisons between the cohesive and non-cohesive models at a scale of one micrometer reveal negligible differences in the elastic modulus. This observation can be attributed to the fact that the composite’s elastic modulus is assessed under relatively low deformation, where insufficient strain prevents interface separation. Consequently, it is reasonable to conclude that the non-bond (\textit{van der Waals}) strength-induced cohesive interaction does not significantly influence Young's modulus. Although a monotonic increase in the modulus is observed while increasing fiber volume fraction, the effect of the interfacial cohesive interaction on the modulus remains insignificant. Thus, while the scaling effect substantially influences the UTS of composites as the scale of dimension decreases from one micrometer to one nanometer, its impact on the homogenized elastic modulus is relatively minor.

	\section{Conclusion}
	\label{Sec5: Conclusion}
	
	This study presents a combined hierarchical molecular dynamics and phase-field multiscale modeling approach to investigate the fracture behavior of nanocomposites from the nano- to the microscale. We investigated the scaling effects and fracture responses of polymer composites reinforced by graphene and carbon nitride nanomembranes. By integrating molecular dynamics with an anisotropic star-convex cohesive phase-field model, this approach captures the intrinsic \textit{van der Waals} forces driving cohesive interactions, from nanoscale interactions to macroscale fracture behavior. The simulation results revealed a remarkable scaling effect: the tensile strength of the composites increased nonlinearly with the filler thickness, especially for graphene-reinforced composites with filler thicknesses ranging from 1 nm to 10 nm. Our multiscale modeling results indicate that for thin nanosheets, the fiber-matrix interface strength plays a dominant role in enhancing fracture resistance, whereas for thicker fillers, fracture resistance is influenced by the reinforcement distribution and density, leading to a softening effect.
	
	Furthermore, the star-convex cohesive phase-field model successfully splits the asymmetric fracture response between tension and compression, allowing the simulation of damage evolution for complex RVEs. This approach highlights the critical influence of fiber-induced cohesive interactions on the mechanical resilience of composites, with notable increases in Young's modulus and toughness occurring with increasing reinforcement volume and size. These findings suggest that the size and reinforcement proportion directly dictate the fracture mechanics of the composites, impacting their toughness and resistance to crack propagation, especially in global softening regions where fiber-matrix interfaces concentrate stress.
	
	Despite the exciting results obtained, this approach currently omits temperature-dependent fracture behavior, and MD simulations rely on empirical interatomic potentials. Future work could incorporate temperature-coupled cohesive fracture models with machine learning interatomic potentials 
	, thereby enabling the investigation of fracture mechanics under thermal variations in heterostructures and nanocomposites. Furthermore, the proposed model does not take into account the flexible nature of nanosheets, which requires the construction of more complex RVEs. The transversely isotropic mechanical nature of layered two-dimensional materials is also not considered, which can be investigated in future studies.

	\section{Acknowledgements}
	\label{Acknowledgement} 
	B. Mortazavi, X. Zhuang, and N. Valizadeh appreciate the funding by the DFG under Germany’s Excellence Strategy within the Cluster of Excellence PhoenixD (EXC 2122, Project ID: 390833453). M. Liu acknowledges the funding support from the China Scholarship Council (CSC). The authors acknowledge the valuable discussions from Fadi Aldakheel.

	\appendix
	\section{Tersoff potential expansion}
	\label{AppendixA}
	
	\[
	{\psi _{{\rm{Tersoff}}}} = f_{C} \, ( {\underbrace {\mathcal{A} \, {e^{ - {\lambda _1}r}}}_{{f_R}} - \underbrace {{b_{ij}} \, \mathcal{B} \, {e^{ - {\lambda _2}r}}}_{{f_A}}})
	\]
	The lower indices $i$ and $j$ represent atomic labels rather than the indices notation used in continuum mechanics.
	
	\[{f_C} = \left\{ {\begin{array}{*{20}{l}}
			{1,}&{r < {R_{ij}}}\\
			{\frac{1}{2}\left[ {1 + \cos \left( {\pi {\textstyle{{r - {R_{ij}}} \over {{S_{ij}} - {R_{ij}}}}}} \right)} \right],}&{{R_{ij}} < r < {S_{ij}}}\\
			{0,}&{r > {S_{ij}}}
	\end{array}} \right.\]
	
	\noindent $R_i$ denotes the interaction cutoff for atom $i$. $R_{ij}$ represents the bond cutoff between atoms $i$ and $j$. 
	\[R_{ij}=(R_{i} \, R_{j})^{1/2}, \quad S_{ij}=(S_{i} \, S_{j})^{1/2}, \quad {\lambda _1} = (\lambda _{1 \, i} + \lambda _{1 \, j})/2, \quad {\lambda _{2}} = (\lambda _{2 \, i} + \lambda _{2 \, j})/2\]
	
	\noindent The index $k$ denotes the third atom out of the plane defined by atoms $i$ and $j$. The angle $\theta_{ijk}$ represents the three-body angle among atoms $i$, $j$, and $k$, which contributes to the attractive pairwise interaction $f_{A}$. The parameter $m$ is dependent upon the type of atoms involved. For example, in the case of graphene, only the atom C is considered. In contrast, for the case of carbon-nitride, both the atom C and the atom N must be identified.
	
	\[
	{b_{ij}} = {\left( {1 + \beta ^{m} \, \varsigma _{ij}^m} \right)^{ - 1/2m}}, \quad
	\varsigma _{ij}^n(\theta _{ijk}) = \sum\limits_{k \ne i,j} {{f_C} \, g({\theta _{ijk}}) \, e^{\lambda _{3}^{3} \, [r-r_{ik}^3]}} , \quad g({\theta _{ijk}}) = \left( {1 + \frac{{c_i^2}}{{d_i^2}} - \frac{{c_i^2}}{{d_i^2 + {{(\cos ({\theta _{ijk}}) - {h_i})}^2}}}} \right)
	\]
	
	\noindent Here, detailed parameters of GN were listed as follows:
	
	\begin{table}[h]
		\begin{tabular}{llll}
			\hline
			$\mathcal{A}$ ($eV$)        & 1393.6  & $\mathcal{B}$ ($eV$)               & 430.0   \\
			$\lambda _1$ (1/\text{\AA}) & 3.4879  & $\lambda _1$ (1/\text{\AA})        & 2.2119  \\
			$\lambda _3$ (1/\text{\AA}) & 0.0     & $m$ (-)                    & 0.72751 \\
			$c$ (-)             & 38049.0 & $\beta \, (10^{-7})$ & 1.5724  \\
			$d$ (-)             & 4.3484  & $h$ (-)                    & -0.930  \\
			$R$ (\text{\AA})         & 2.05    & $S$ (\text{\AA})                & 0.05    \\ \hline
		\end{tabular}
		\centering
		\caption{The atomic coefficients of the \textit{Tersoff} potential for modeling the mechanical response of graphene.}
	\end{table}
	
	\section{Cohesive potentials}
	\label{AppendixB}
\begin{figure*}[!h]
		\centering
		\includegraphics[width=19cm]{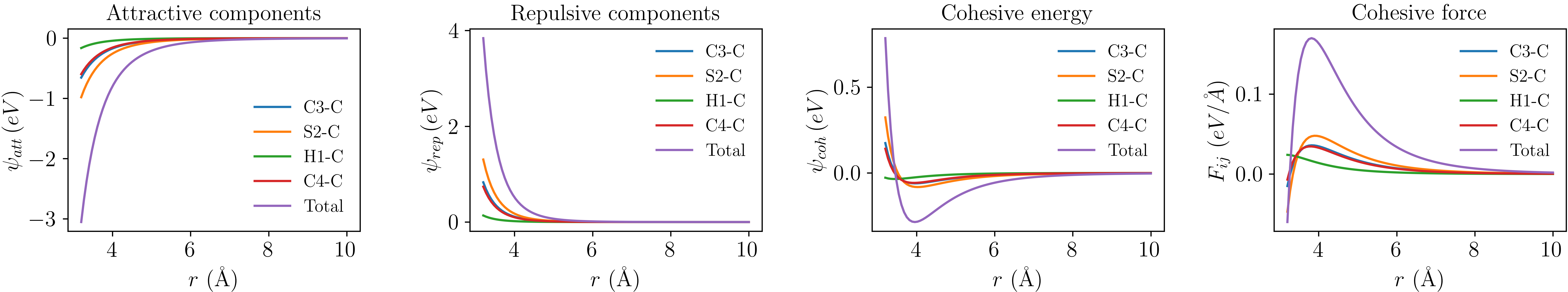}
		\caption{The detailed potential components and cohesive force components for the GN-P3HT cohesive model.}
		\label{fig10}
	\end{figure*}

	The cohesive potential $\psi_{\rm{coh}}$ is decomposed as the attractive potential $\psi_{\text{att}}$ and repulsive potential $\psi_{\text{rep}}$, which are compared between the carbon (C2, C3, C4), sulfur (S2), and hydrogen (H1) atoms and carbon (C) atom from graphene. C2 and C3 share the same bond coefficients. The attractive and repulsive potentials are defined as $\psi_{\text{att}} = - 3 \, \epsilon \, {{\left( {\frac{{{E _0}}}{r}} \right)}^6}$ and $\psi_{\text{rep}} = 2 \, \epsilon \, {{\left( {\frac{{{E _0}}}{r}} \right)}^9} $. The detailed atomic coefficients for the traction-separation model are listed in \hyperref[AppendixTable1]{Table B.3}.
	\begin{table}[]
		\centering
		\begin{tabular}{clll}
			\hline
			\multicolumn{1}{l}{}      & Atoms & $\epsilon$ ($eV$) & $E_0$ (\AA) \\ \hline
			\multirow{4}{*}{Matrix}   & C3/C2 & 0.068           & 3.915                 \\
			& S2    & 0.125           & 4.047                 \\
			& H1    & 0.023           & 2.878                 \\
			& C4    & 0.062           & 3.854                 \\ \hline
			\multicolumn{1}{l}{Fiber} & C     & 0.054           & 4.010                 \\ \hline
		\end{tabular}
		\caption{The atomic coefficients for modeling the interfacial interaction between fiber and matrix.}
		\label{AppendixTable1}
	\end{table}
	\newpage
	By following the Waldman-Hagler rules \cite{waldman1993new}, the mixed interfacial cohesive coefficients are calculated. The detailed potential energies and their corresponding cohesive forces are depicted in the following graphs, see \hyperref[fig10]{Fig.B.11}. The cohesive force $F_{ij}$ between two atoms $i$ and $j$ is calculated by the first derivative of the cohesive potential $\psi_{{\rm{coh}}}$ respect to the actual distance $r_{ij}$ of the atoms as $F_{ij} =  - \frac{\partial \psi _{\rm{coh}}}{\partial r_{ij}}$. 
	Therefore, the interaction force between graphene and matrix can be decomposed into C2-C, C3-C, S2-C, H1-C, and C4-C bonds. Hence, their detailed cohesive contributions are depicted in the following.
	
	The attractive potential $\psi_{\text{att}}$ and repulsive potential $\psi_{\text{rep}}$ are compared by Lennard-Jones 12-6 and Lennard-Jones 9-6 potential expressions.
	\begin{figure}[!h]
		\centering
		\includegraphics[width=16.5 cm]{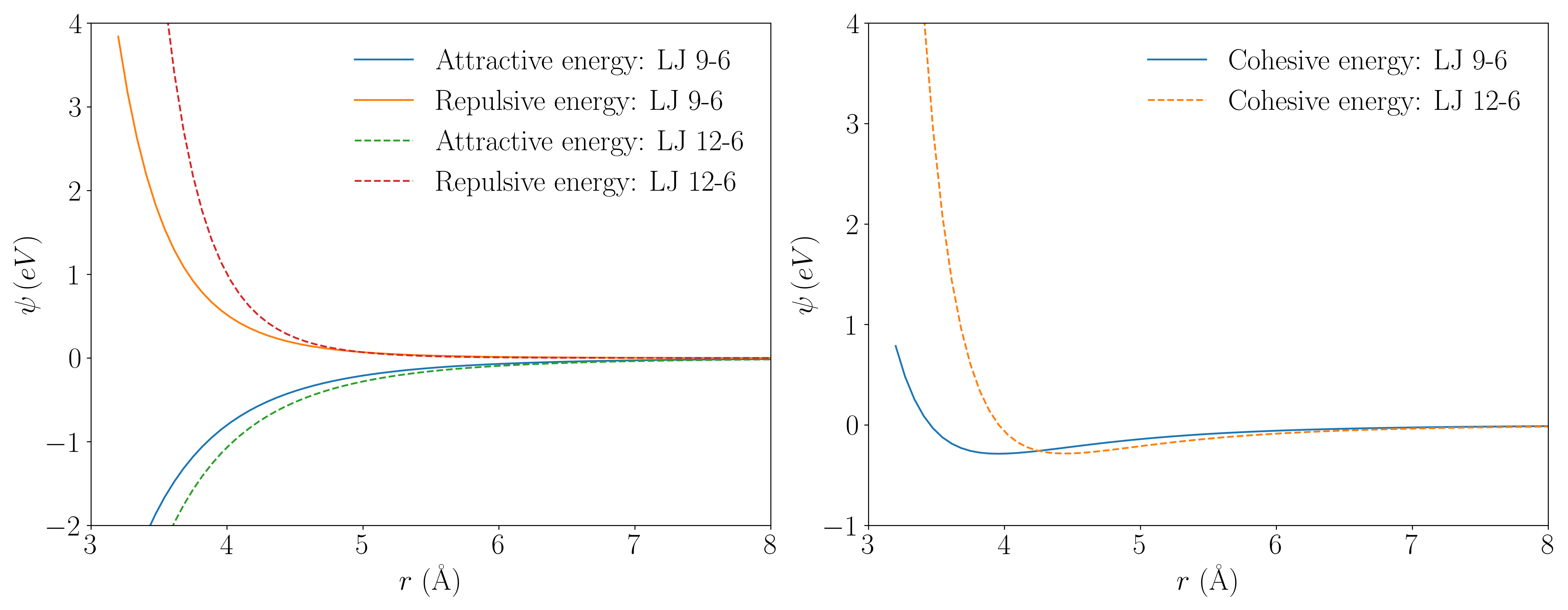}
		\caption{Comparison between the Lennard-Jones 9-6 and Lennard-Jones 12-6 potentials.}
		\label{fig11}
	\end{figure}
	
	\section{Derivations}
	\label{AppendixC}
	\subsection{Derivations for $g'(d)$}
	\noindent  The detailed derivations of the degradation function $g(d)$ with a polynomial order of $p=2$ with respect to the phase-field variable $d$ are expressed as follows:
	\begin{equation}
		\begin{split}
			g'(d) &= - \frac{(1 - d)^2 (-2 \, (1 - d) + a_1 \, d \, (a_2 + 2 \, a_3 \, d) + a_1 \, (1 + a_2 \, d + a_3 \, d^2))}{((1 - d)^2 + a_1 \, d \, (1 + a_2 \, d + a_3 \, d^2))^2}  - \frac{2 \, (1 - d)}{(1 - d)^2 + a_1 \, d \, (1 + a_2 \, d + a_3 \, d^2)} \\
			g'(d) &= \frac{a_1 \, (d - 1) (1 + d + 2 \, a_2 \, d + 3 \, a_3 \, d^2 - a_3 \, d^3)}{(1 + (a_1 - 2) \, d + (1 + a_1 \, a_2) d^2 + a_1 \, a_3 d^3)^2}.
		\end{split}
	\end{equation}
	
	\subsection{Derivations for $a_1$}
	\noindent The derivations to obtain the variable $a_1$ in one-dimensional case are expressed as follows:
	\label{AppendixC-1}
	\begin{equation}
		\begin{split}
			\frac{{\partial {\psi _{{\text{in}}}}}}{{\partial d}} &= \frac{{\partial {\psi _{{\text{bulk}}}} }}{{\partial d}} + \frac{\partial \psi_{{\rm{frac}}}}{\partial d},\\
			\frac{{\partial {\psi _{{\text{in}}}}}}{{\partial d}}  & = \frac{{\partial g(d)}}{{\partial d}}{\psi _e} + \frac{\partial }{{\partial d}}\left[ {{G_c} \, \gamma^i (d,\nabla d)} \right],\\
			\frac{{\partial {\psi _{{\text{in}}}}}}{{\partial d}}  & = \frac{{\partial g(d)}}{{\partial d}}{\psi _e} + \frac{\partial }{{\partial d}}\left[ {\frac{{{G_c}}}{{{c_0}}}\left( {\frac{1}{{{l_f}}}\omega (d) + {l_f}\nabla d \cdot \nabla d} \right)} \right],\\
			\frac{{\partial {\psi _{{\text{in}}}}}}{{\partial d}}  & = g'(d){\psi _e} + \frac{{{G_c} \, \omega '(d)}}{{{c_0} \, {l_f}}}.
		\end{split}
	\end{equation}
	The detailed expansion for $\omega'(d)$, $\psi _e$, and $g'(d)$ are expressed as:
	\begin{equation}
		\left\{ 
		\begin{split}
			&\omega '(d) = 2 \, (1 - d),\\
			&{\psi _e} = \frac{1}{2} \, {\sigma _0} \, \varepsilon ,\quad \text{with} \quad \varepsilon  = \frac{{{\sigma _0}}}{E} \to {\psi _e} = \frac{{\sigma _0^2}}{{2 \, E}},\\
			&g'(d = 0) =  - {a_1}.
		\end{split}
		\right.\\
	\end{equation}
	When $d=0$, the partial derivative of $\frac{\partial \psi_{{\rm{in}}}}{\partial d} = 0$ can be expressed as: 
	\begin{equation}
		\begin{split}
			\frac{{\partial {\psi _{{\text{in}}}}}}{{\partial d}} &= {\left. {\left( {g'(d)\frac{{\sigma _0^2}}{{2 \, E}} + \frac{{2 \, {G_c} \, (1 - d)}}{{{c_0} \, {l_f}}}} \right)} \right|_{d = 0}},\\
			\frac{{\partial {\psi _{{\text{in}}}}}}{{\partial d}}  &=  - {a_1}\frac{{\sigma _0^2}}{{2 \, E}} + \frac{{2 \, {G_c}}}{{{c_0} \, {l_f}}} = 0.
		\end{split}
	\end{equation}
	To the end, the variable $a_1$ can be obtained as:
	\begin{equation}
		{a_1} = \frac{{4 \, E \, {G_c}}}{{{c_0} \, {l_f} \, \sigma _0^2}} = \frac{{4 \, {l_{ch}}}}{{{c_0} \, {l_f}}}, \quad {\text{with }} \quad {l_{ch}} = \frac{{E \, {G_c}}}{{\sigma _0^2}}.
	\end{equation}
	
	\newpage	
	\section*{References}
	\bibliography{Article}
	
\end{document}